\documentclass[]{aa}     

\usepackage[varg]{txfonts}
\usepackage[english]{babel}
\usepackage{graphicx}   
\usepackage{epstopdf}
\usepackage{amsmath}
\usepackage{amssymb}
\usepackage{float}
\usepackage{float}
\usepackage{color}
\usepackage{enumitem}
\usepackage{soul} 

\newcommand{\mx}{\mathbf{x}}
\newcommand{\tOm}{\widetilde{\mathbf{\Omega}}\mathbf{_p}}
\newcommand{\tOmd}{\dot{\widetilde{\mathbf{\Omega}}}\mathbf{_p}}

\begin{document}

\title{From Manifolds to Lagrangian Coherent Structures in galactic bar models}

\author{P. S\'anchez-Mart\'{i}n\inst{\ref{inst1}}\and J. J. Masdemont\inst{\ref{inst2}}\and M. Romero-G\'{o}mez\inst{\ref{inst3}}
}

\institute{GTM - Grup de recerca en Tecnologies M\`edia, La Salle, Universitat Ramon Llull, Quatre Camins 2, E08022 Barcelona, Spain, \email{psanchez@salleurl.edu}\label{inst1} \and IEEC-UPC i Dept. de Matem\`{a}tiques, Universitat Polit\`{e}cnica de Catalunya, Diagonal 647 (ETSEIB), E08028 Barcelona, Spain, \email{josep.masdemont@upc.edu}\label{inst2} \and Dept. d'Astronomia i Meteorologia, Institut de Ci\`{e}ncies del Cosmos, Universitat de Barcelona, IEEC, Mart\'{i} i Franqu\`{e}s 1, E08028 Barcelona, Spain, \email{mromero@fqa.ub.edu}\label{inst3}
}

\date{}

\abstract{
We study the dynamics near the unstable Lagrangian points in galactic bar
models using dynamical system tools in order to determine the global morphology
of a barred galaxy. We aim at the case of non-autonomous models, in particular
with secular evolution, by allowing the bar pattern speed to decrease with
time. We extend the concept of manifolds widely used in the autonomous problem
to the Lagrangian Coherent Structures (LCS), widely used in fluid dynamics,
which behave similar to the invariant manifolds driving the motion.  After
adapting the LCS computation code to the galactic dynamics problem, we apply it
to both the autonomous and non-autonomous problems, relating the results
with the manifolds and identifying the objects that best describe the motion in
the non-autonomous case.  We see that the strainlines coincide with the first
intersection of the stable manifold when applied to the autonomous case, while,
when the secular model is used, the strainlines still show the regions of
maximal repulsion associated to both the corresponding stable manifolds and
regions with a steep change of energy. The global morphology of the galaxy
predicted by the autonomous problem remains unchanged.
}


\maketitle

\section{Introduction} \label{intro}

Lagrangian Coherent Structures (LCS), introduced by \citet[][]{Haller00}, have
been proposed as dynamical replacements for invariant manifolds in the study of
the dynamics of non-autonomous systems. They behave as hypersurfaces with
maximally attracting or repelling properties and organizing the evolution of
the system in a similar way as invariant manifolds do in autonomous problems.

LCS have been succesfully employed in the study of fluid dynamics 
problems~\citep[e.g.][]{Haller12,Haller13}, or in the elliptic restricted
three-body problem (ER3BP)~\citep[see][]{Gawlik}. The exact concept of what a
LCS is, is still evolving. \citet[][]{Shadden05,Lekien07,Gawlik}, among others,
consider them as ridges of the values of finite-time Lyapunov exponents (FTLE)
of the flow, whereas~\citet{Haller12} and therein, \citet{Haller14},
characterize them as critical lines of the averaged material shear, which is an
auxiliary autonomous functional also derived from the flow.

We seek to apply the theory of Lagrangian Coherent Structures to the study of
the non-autonomous version of the precessing galactic bar model~\citep{Warps}.
This model is a Hamiltonian system which in its autonomous version has been
studied by means of invariant manifolds. Our goal in this paper is to  compare
the results provided by the invariant manifolds with those of the LCS in the
autonomous version, and furthermore, to study a simple non-autonomous version
of this problem.  We have developed software to implement the characterization
of LCS given by Haller in two dimensions~\citep{Haller14}, such as stretch and
strainlines. In the work it is to determine LCS in both autonomous and
non-autonomous versions of a precesing galactic bar potential.


LCS are a quantitative characterization of chaotic transport, based on the fast dynamical indicator FTLE.
In the literature there exist analogous methods to detect chaotic and ordered orbits of a
dynamical system, such as the smaller alignment index (SALI) \citep[][]{Skokos01}, or the Mean Exponential Growth factor of Nearby
Orbits (MEGNO) \citep{Cincotta}.
The main idea of the SALI method is to study the evolution in time
of two different deviation vectors. In this case, if SALI tends to zero the
orbit is chaotic, whereas if it tends to a positive non-zero value the orbit is
ordered. The eficiency of the SALI method has been widely proved in well known
problems \citep[][]{Skokos04} and in barred galaxies models \citep[][]{Skokos08}.
The MEGNO method is a refinement of the Lyapunov Characteristic Number (LCN):
the latter studies the mean exponential rate of divergence of nearby orbits
by integrating them over a long time span, while MEGNO is a time-weighted version
of the LCN which can be found by integrating the perturbed orbits over a
shorter time span and still detects the regularity or chaocity of the system,
and estimates its hyperbolicity in the latter case. MEGNO has been applied to
the study of the dynamics of elliptical galaxies \citep[][]{Cincotta08}, and 
of planetary systems \citep[][]{GBM01}.

The paper is organized as follows: In Sect.~\ref{sec2:LCS} we review the theory
concerning Lagrangian Coherent Structures (LCS) and their correspondence with
finite-time Lyapunov exponents (FTLE). Sect.~\ref{sec3:compuLCS} discusses in
detail how to compute LCS in practice.  The application of LCS to the galactic
model, and their relation with the classical invariant manifolds is shown in
Sect.~\ref{sec4:LCSprec}. Sect.~\ref{sec5:conclu} presents the conclusions of
the paper, and outlines directions for future work in the topic.



\section{Lagrangian Coherent Structures} \label{sec2:LCS}

The starting point to study the local and global structure of a time-dependent
flow is the Jacobian given by the variational flow. Let us consider a dynamical
system of the form,
\begin{equation}\label{eqn:dyn} \mathbf{\dot{x}} =
\mathbf{v}(\mathbf{x},t), \quad \mathbf{x} \in U \subset \mathbb{R}^n, \quad
  t\in [a,b], 
\end{equation} 
where $U$ denotes an open, bounded subset of
$\mathbb{R}^n$, the time $t$ varying over the finite interval $[a,b]$, and
$\mathbf{v}: U \times [a,b] \to \mathbb{R}^n$ is a sufficiently smooth vector field. 
For $a \leq t_0<t \leq b$, we define the flow map 
\begin{equation}
\mathbf{F}_{t_0}^t(\mathbf{x}_0) := \mathbf{x}(t), 
\end{equation} 
where $\mathbf{x}(t)$ is the solution of equation~\eqref{eqn:dyn} such that
$\mathbf{x}(t_0)=\mathbf{x}_0$.

For a fixed time $t$, the Jacobian $\nabla\mathbf{F_{t_0}^t}(\mathbf{x}_0)$ provides 
a linearization of the
variation of the flow $\mathbf{F_{t_0}^t}$ with respect to the initial
condition $\mx_0$. Its Singular Value Decomposition (SVD) points out the
directions where the flow is maximally spread (maximally compressed), with
the rate of expansion (compression) given by the singular values of the
Jacobian. For instance, the maximal expansion rate of the flow around $\mx_0$
is the first singular value of its Jacobian, which is the Euclidean
norm\footnote{The Euclidean or spectral norm of a matrix $A$ is
$||A||=\underset{v\neq0}{\sup}\frac{||Av||}{||v||}$, or equivalently, the maximum
singular value of $A$.} of the Jacobian
$||\nabla\mathbf{F_{t_0}^t}(\mathbf{x}_0)||$, and the direction where this
maximal expansion happens is given by its associated right-singular vector of
the SVD.

The SVD of the Jacobian $\nabla\mathbf{F_{t_0}^t}(\mathbf{x}_0)$ is equivalent
to the diagonalization of the Cauchy-Green, or strain, tensor field, widely
used in Mechanics:
\begin{equation}
 \mathbf{C}_{t_0}^t = (\nabla\mathbf{F_{t_0}^t}(\mathbf{x}_0))^T \, \nabla\mathbf{F_{t_0}^t}(\mathbf{x}_0),
\end{equation}
where $^T$ stands for matrix transposition. 

The eigenvalues of the Cauchy-Green tensor are the \emph{squares} of the
singular values of the Jacobian, and the eigenvectors of the Cauchy-Green
tensor, $\xi_1,\, \xi_2, \,\dots, \xi_n$, are the right-singular vectors of the
Jacobian. In both instances an orthonormal basis of vectors is adopted, but,
unfortunately, opposite sorting conventions are followed: It is usual to
sort the singular values from largest to smallest,
$||\nabla\mathbf{F_{t_0}^t}(\mathbf{x}_0)|| = \sigma_1 \ge \sigma_2 \ge \ldots
\ge \sigma_n$, while the eigenvalues of the Cauchy-Green tensor are usually
labelled in increasing order, $\lambda_1 \le \lambda_2 \le \ldots \le
\lambda_n$. See~\citet{Golub12} for a mathematical discussion of these
concepts.

The concept of a Lagrangian Coherent Structure is somewhat recent and still 
work in progress. Several versions with slight differences can be found in the literature.
We shall use in this work the definition of LCS given by Haller, and the following
characterization of LCS in two dimensions, based on variational theory,
from~\citet{Haller12}: 

A line $\mathcal{M}(t_0)$ in the two dimensional spatial domain of the dynamical system is a repelling LCS (or strainline) 
for the system over the time interval $[t_0,t]$ if and only if, for every point $\mx_0 \in \mathcal{M}(t_0)$ the following conditions hold,
 \begin{enumerate}
  \item $\lambda_1(\mx_0) \neq \lambda_2(\mx_0) > 1$;\label{theo:replcs1}
  \item $\langle \xi_2(\mx_0),\nabla^2\lambda_2(\mx_0)\xi_2(\mx_0) \rangle < 0$;\label{theo:replcs2}
  \item $\xi_2(\mx_0) \perp \mathcal{M}(t_0)$;\label{theo:replcs3}
  \item $\langle \nabla \lambda_2(\mx_0), \xi_2(\mx_0) \rangle =0 $,\label{theo:replcs4}
 \end{enumerate}
where $\lambda_1$ and $\lambda_2$, are respectively the smallest and largest 
eigenvalues of the Cauchy-Green tensor of the flow, and $\xi_1,\, \xi_2$ are
the corresponding eigenvector fields.
Note that condition~\eqref{theo:replcs1} ensures that the stretching rate of
the flow (i.e. the rate at which particular solutions of the system separate
when integrated over the time interval $[t_0,t]$) is greater along the normal
direction that along the tangential direction. Conditions~\eqref{theo:replcs3}
and \eqref{theo:replcs4} assure that the normal stretch rate of the flow along
the strainline is a local extremum relative to close material lines,
while condition~\eqref{theo:replcs2} assures this extremum as a strict local maximum.
 
The characterization of an attracting LCS (or stretchline) is analogous to that
of a strainline, but replacing the second eigenvalue and vector of the
Cauchy-Green tensor by the first ones, and reversing the inequality in
condition~\eqref{theo:replcs2} above. 

\begin{figure}
\centering
    \includegraphics[width=0.5\textwidth]{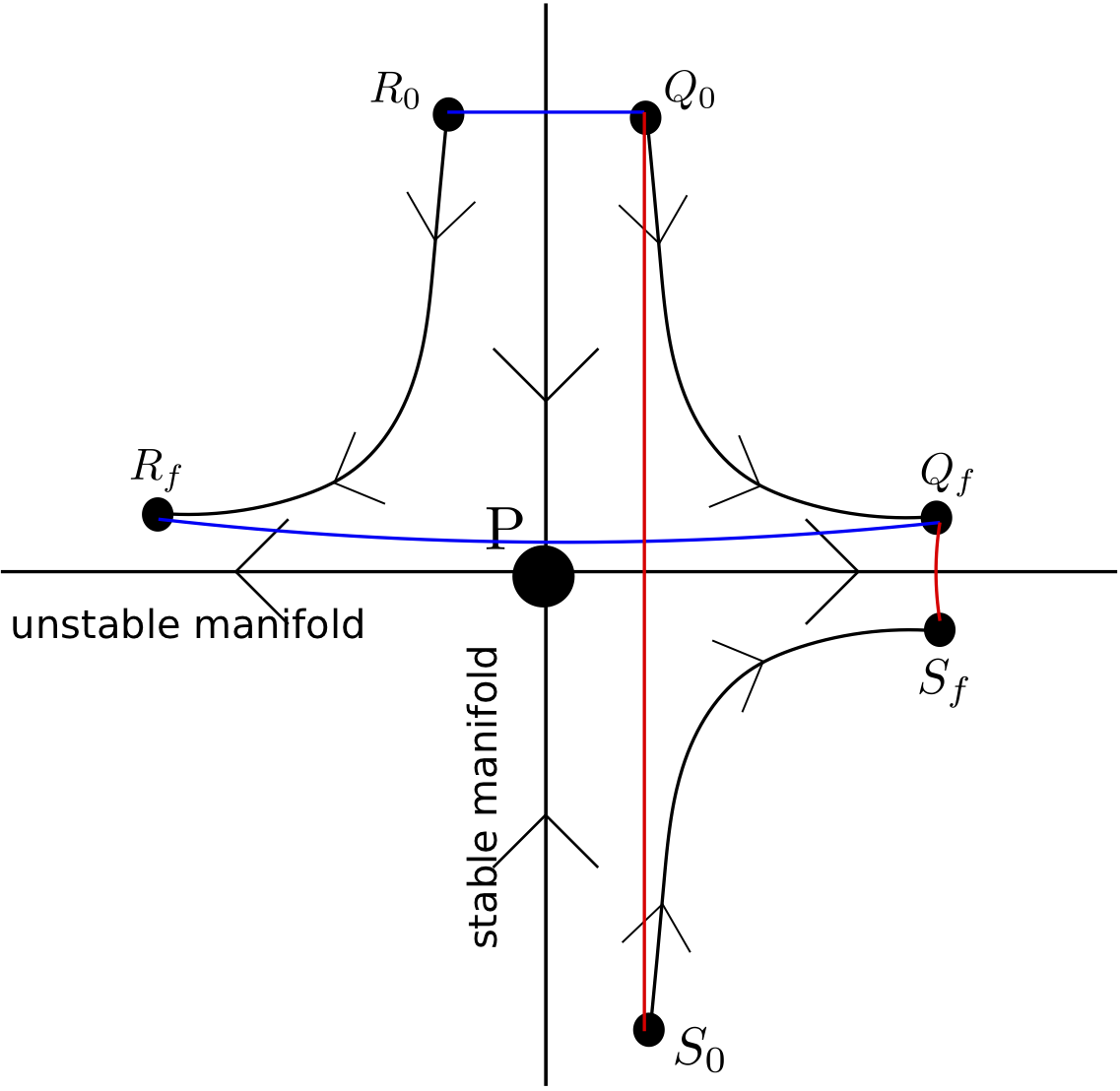} 
    \caption{Neighbourhood of a hyperbolic equilibrium point in a 2-dimensional, autonomous, dynamical system:
The blue segment of initial conditions from $Q_0$ to $R_0$, transverse to the stable invariant manifold, will undergo
great stretch when the system is integrated from time $t_0$ to $t_f$ and it transforms into the curve from $Q_f$ to $R_f$. Conversely, 
the red segment of initial conditions from $Q_0$ to $S_0$ will undergo great compression when the system is integrated 
and it transforms into the curve from $Q_f$ to $S_f$. Any segment of initial conditions not intersecting the 
invariant manifolds will undergo lesser stretches or strains.}
\label{fig:lcsvsinv}
\end{figure}

For a dynamical system close to being autonomous, in the neighbourhood of a 
hyperbolic equilibrium point the repelling LCS are the
indicator analogue of stable invariant manifolds, while attracting LCS are 
the analogue of unstable invariant manifolds. The reason for this is illustrated
by Fig.~\ref{fig:lcsvsinv}: a segment of initial conditions transverse to the stable
manifold will stretch to a great extent when integrated beyond the equilibrium point,
while a segment of initial conditions transverse to the unstable manifold will 
be greatly compressed by the flow. Both these stretchings and compressions 
are the greatest taking place in the neighbourhood of the equilibrium point.

This definition of attracting and repelling LCS from Haller et al extends to higher dimensions:
one just integrates the system forward
in time and considers the largest and the smallest eigenvalues of the
Cauchy-Green tensor.
According to~\citet{Haller13} a \emph{strain-surface} or \emph{repelling LCS}
is a codimension one hypersurface in the spatial domain of the dynamical system
such that at initial time $t_0$ it is everywhere normal to the eigenvector
field of the largest eigenvalue of the Cauchy-Green tensor. A
\emph{stretch-surface} or \emph{attracting LCS} is a codimension one
hypersurface in the spatial domain of the dynamical system such that at initial
time $t_0$ it is everywhere normal to the eigenvector field of the smallest
eigenvalue of the Cauchy-Green tensor.
Roughly speaking, a repelling LCS is an hypersurface that over the taken integration 
time interval is pointwise more repelling than any nearby hypersurface. On
the contrary, an attracting LCS maximizes pointwise attraction when integrating
the dynamical system among nearby hypersurfaces.

Let us point out that, in any dimension, the definition of LCS that we are using
differs from the alternative one from \citet{Shadden05}, which is the one used in \citet{Gawlik} to study the dynamics of a non-autonomous elliptic restricted three body problem by determining its LCS in the role played by invariant manifolds in the autonomous case. According to \citet{Shadden05}, a repelling LCS
is a ridge of the Finite Time Lyapunov Exponent (FTLE) field of the flow $\mathbf{F}_{t_0}^t(\mathbf{x}_0)$,
and an attracting LCS is defined in the same way, except that the dynamical 
system is integrated backwards in time for the computation of the FTLE field.

The inability of FTLE ridges, even in some autonomous flows, to
completely explain the flow pattern in certain situations led Haller
to propose his alternative definition. In our 2-dimensional setting,
the conditions (2) and (4) of the last characterization of repelling
LCS are satisfied by ridges of the FTLE field, but conditions (1) and (3)
not necessarily. \citet{Haller11}
presents examples of repelling LCS which are not FTLE ridges, and of FTLE
ridges that are not repelling LCS.

.png\section{The computation of Lagrangian Coherent Structures} \label{sec3:compuLCS}


It is convenient to relax conditions~(\ref{theo:replcs2})
and~(\ref{theo:replcs4}) characterizing a repelling LCS in the previous section
because of numerical computation reasons. Condition~\eqref{theo:replcs2} is
problematic because the eigenvalue $\lambda_2(\mx_0)$ may be locally constant
over part of the domain and in this case the numerical computation of strainlines
becomes unstable.
It is recommended in such cases to allow the LCS to have non-zero thickness.
According to~\citet{Haller12}, condition~\eqref{theo:replcs4} is often numerically sensitive and it is advisable to replace this local
condition by its average along the strainline.  Relaxing
condition~\eqref{theo:replcs4} in this way is consistent with numerical and
laboratory observations of tracer mixing in two-dimensional fluid
flows~\citep{Haller12}.

So, recalling that $\xi_1, \,\xi_2$ form an orthonormal basis of the plane, the
alternative conditions to~\eqref{theo:replcs1} -~\eqref{theo:replcs4} are,
\begin{enumerate}[label=(\roman*)]
 \item $\lambda_1(\mx_0) \neq \lambda_2(\mx_0) > 1$;\label{theo:replcs1p}
 \item $\langle \xi_2(\mx_0),\nabla^2\lambda_2(\mx_0)\xi_2(\mx_0) \rangle \leq 0$; \label{theo:replcs2p}
 \item $\xi_1(\mx_0) \, || \, \mathcal{M}(t_0)$;\label{theo:replcs3p}
 \item $\bar\lambda_2(\gamma)$, the average of $\lambda_2$ over a curve $\gamma$, is maximal on $\mathcal{M}(t_0)$ among all nearby curves $\gamma$ satisfying $\dot{\gamma}(t_0) \, || \, \xi_1(\mx_0)$. \label{theo:replcs4p}
\end{enumerate}

\begin{figure}
\centering
    \includegraphics[width=0.5\textwidth]{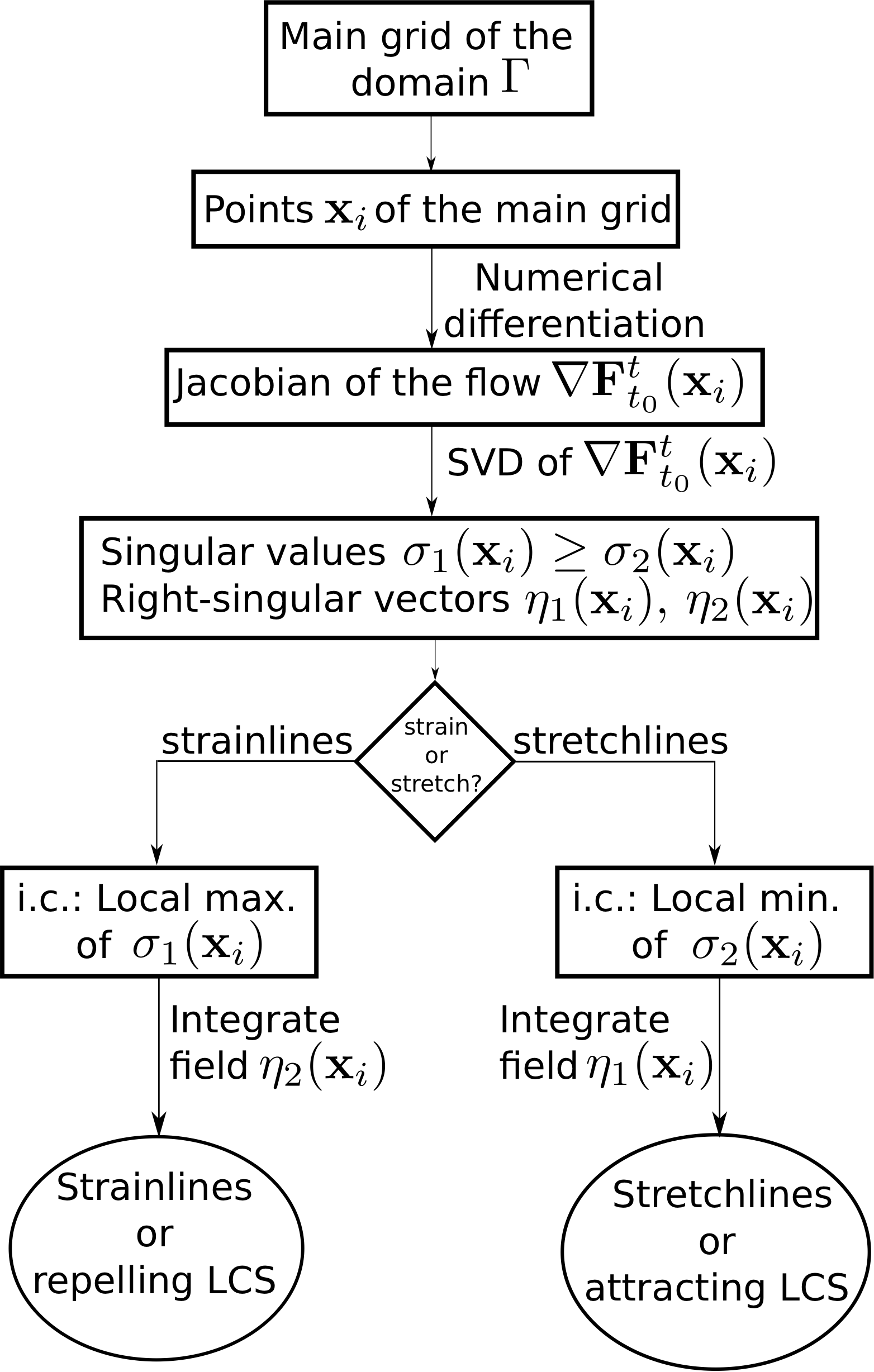} 
    \caption{Flowchart of the algorithm followed for the computation of LCS. Note that it requires integration of two dynamical systems. First, the flow of the original dynamical system on a grid must be computed, in order to estimate its Jacobian. After that, we take as initial condition (i.c.) local extrema of the singular values and integrate the field given by the corresponding right-singular vectors.}
  \label{fig:flowchart}
\end{figure}

In order to create our software package for the computation of LCS, we follow
the algorithm given in~\citet{Haller14}, which implements a characterization
shown in~\citet{Haller14shear} to be equivalent to
conditions~\eqref{theo:replcs1} -~\eqref{theo:replcs4} and based on the
integration of the autonomous vector fields given by the eigenvectors of the
Cauchy-Green tensor (i.e. the left-singular vectors of the Jacobian
$\nabla\mathbf{F}_{t_0}^t$). We present a schematic flowchart of our implementation in Fig.~\ref{fig:flowchart}

A strainline or repelling LCS is obtained by taking as initial point $\mx_0$ a
local maximum of the largest eigenvalue $\lambda_2$ and integrating from
$\mx_0$ the eigenvector field $\xi_1$ forward and backward in time. 
Analogously, a stretchline or attracting LCS is obtained by taking as initial
point $\mx_0$ a local minimum of the smallest eigenvalue $\lambda_1$ and
integrating from $\mx_0$ the eigenvector field $\xi_2$ forward and backward
in time. 
Let us detail the procedure followed by our software package in order to
compute these strain and stretchlines for the flow $\mathbf{F}_{t_0}^t$ of a
sufficiently smooth dynamical system over a rectangular spatial domain $\Gamma$. 

\begin{figure}
\centering
    \includegraphics[width=0.5\textwidth]{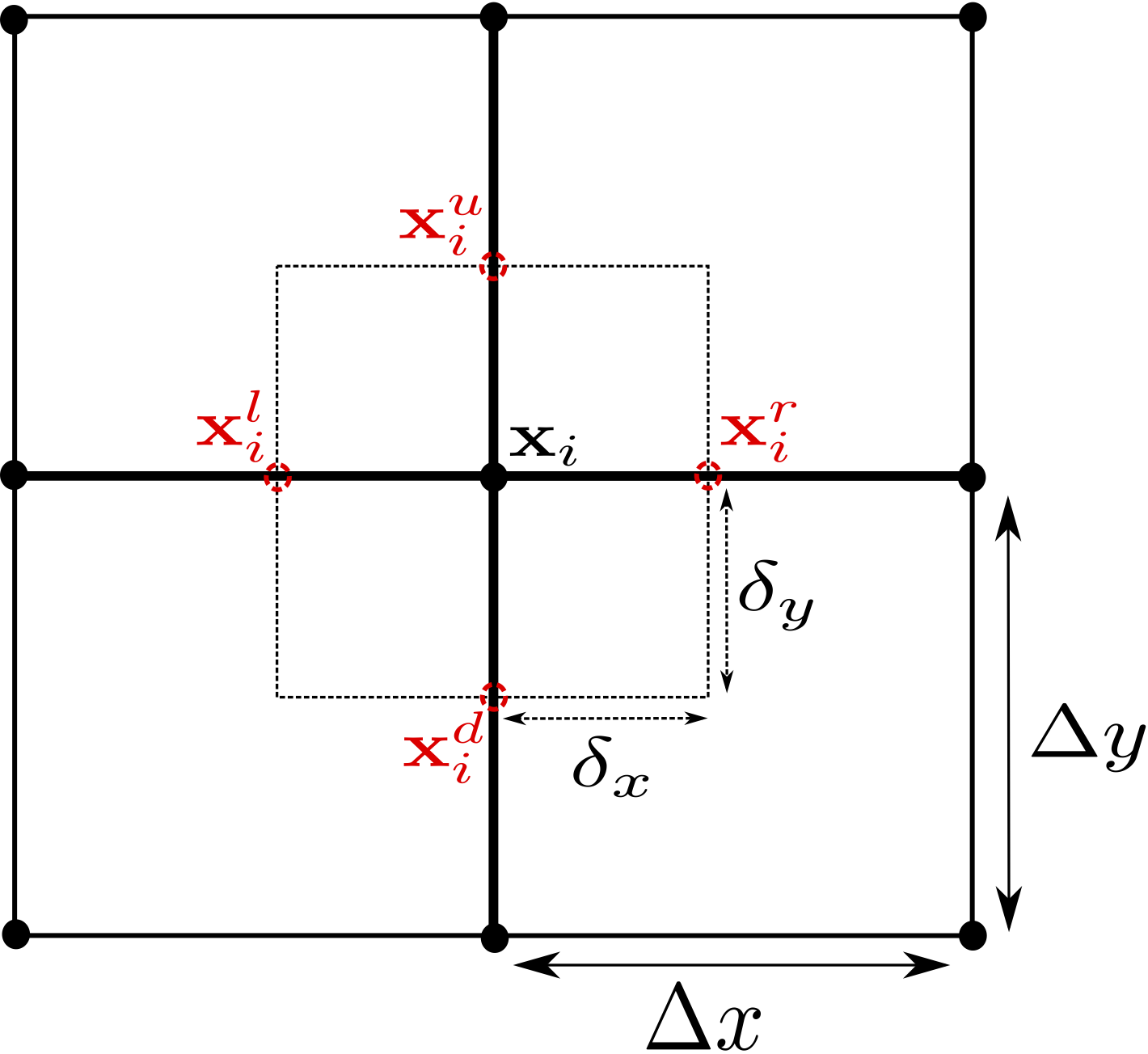} 
    \caption{The four neighboring points (red) to a point $\mx_i$ in the main grid 
     which can be used to compute the Jacobian of the flow 
     $\mathbf{F}_{t_0}^t(\mx_i)$ by finite differences.}
  \label{fig:grid}
\end{figure}

Since the eigenvector fields will be determined numerically in a discrete set
of points, we define a regular rectangular grid, henceforth called the main grid,
covering the domain $\Gamma$ with steps $\Delta x,\, \Delta y$ along the $x, \,
y$ axes, respectively. For each point $\mx_i=(x,y)$ in the main grid we compute the Jacobian of the flow $\mathbf{F}_{t_0}^t$. When variational
equations are available one can eassy take the so called state transition matrix
resulting from the integration with initial conditions $\mx_i=(x,y)$. Otherwise
one can use the same main grid, or even with smaller steps, to approximate it with 
a numerical differentiation formula.
This is, for each point $\mx_i=(x,y)$ in the main grid we can
select four neighboring points $\mx_i^r=(x,y)+(\delta_x,0)$,
$\mx_i^l=(x,y)-(\delta_x,0)$, $\mx_i^u=(x,y)+(0,\delta_y)$,
$\mx_i^d=(x,y)-(0,\delta_y)$, where $\delta_x,\, \delta_y$ define suitable small
displacements (see Fig.~\ref{fig:grid}) and then to implement a centered difference 
skeme,
$$
 \nabla\mathbf{F}_{t_0}^t(\mx_i) \approx \left(\frac{\mathbf{F}_{t_0}^t(\mx_i^r)-\mathbf{F}_{t_0}^t(\mx_i^l)}{2\delta_x},\frac{\mathbf{F}_{t_0}^t(\mx_i^u)-\mathbf{F}_{t_0}^t(\mx_i^d)}{2\delta_y}\right).
$$
The accuracy of this computation is crucial and a high order integrator is advisable
(we use an adaptive Runge-Kutta-Fehlberg method of order 7-8). Special attention must
be paid in the selection of $\delta_{x,y}$ when using the numerical differentiation
approximation.

Once we have computed the Jacobian of the flow $\mathbf{F}_{t_0}^t$ in the points of 
the main grid, the next step is to perform a Singular
Value Decomposition (SVD) on each Jacobian. The purpose of this computation is to
obtain the singular values $\sigma_1(\mx_i) = ||\nabla\mathbf{F}_{t_0}^t|| \ge
\sigma_2(\mx_i)$, and the corresponding right-singular vectors, $\eta_1(\mx_i),
\eta_2(\mx_i)$, for each point of the main grid. We compute directly the SVD of
the Jacobian, rather than the diagonalization of the Cauchý-Green tensor, to
minimize error transmission from the values of the Jacobian to the result of
the computation. In addition, in the cases where the Jacobian is singular, the
SVD yields a singular value very close to zero and positive, while the
diagonalization of the Cauchy-Green tensor, due to numerical rounding errors, 
can produce also an eigenvalue very close to zero but it may have positive or 
negative sign. 

When our domain of the phase state, $U$, has dimension greater than 2, the
computation of 2-dimensional LCS (strain and stretch lines) can still yield a
pretty amount of information about the dynamics of the system by strategically
parameterizing selected surfaces $S$ inside the domain $U$, 
\begin{equation}
\begin{array}{cccc}
 \Psi \, :&  \Gamma & \longrightarrow & S \subset U \\
      &  (\alpha,\beta) & \longmapsto & \mx(\alpha,\beta)
\end{array}
\end{equation}
being again $\Gamma$ a rectangle. The flow $\tilde{\mathbf{F}}_{t_0}^t =
\mathbf{F}_{t_0}^t \circ \Psi$ is now a flow defined on $\Gamma \subset
\mathbf{R}^2$ to which the above computation can be applied. But in this case
special care must be taken to account for the effects of the parameterization $\Psi$.
Now we have $\nabla \tilde{\mathbf{F}}_{t_0}^t=\nabla\mathbf{F}_{t_0}^t \cdot
\nabla \Psi$, and so the Jacobian $\nabla \Psi$ of the parameterization introduces
its own compression or spreading of tangent directions, which do not belong
to the original flow $\mathbf{F}_{t_0}^t$ but it has been artificially inserted by
the parameterization $\Psi$ (for instance, the compression towards the North and South
poles of a sphere introduced by standard spherical coordinates). The solution
to this problem is to apply the SVD to the Jacobian $\nabla\mathbf{F}_{t_0}^t$,
expressed in an orthonormal basis $w_1,w_2$ of the tangent space to the
parameterised surface $S=\Psi(\Gamma)$ at each point $\mx$. This depends only
on the surface $S$ and the point $\mx$, but not on the parameterization $\Psi$.
If $C$ is the base change from this orthonormal basis $w_1,w_2$ to the original
one in our parameterization $\nabla \Psi( \partial_\alpha), \nabla \Psi(
\partial_\beta)$, then we perform the SVD to the matrix
\begin{equation}
(\nabla\mathbf{F}_{t_0}^t)_{|S} (\mx) = \nabla \tilde{\mathbf{F}}_{t_0}^t (\mx) \cdot C
\end{equation}
which is the Jacobian of the flow $\mathbf{F}_{t_0}^t$ restricted to the
surface $S$ using as a departure basis for the tangent space, $T_\mx S$, the
orthonormal basis $w_1,w_2$ which does not introduce distortions to the flow.

Finally, according to the above conditions~\ref{theo:replcs1p} -~\ref{theo:replcs4p},
the strainlines are computed taking the local maxima of the largest singular
value $\sigma_1$ in the main grid as initial condition and integrating the
right-singular vector field $\eta_2$ forward and backward in time. The
stretchlines are computed taking the local minima of the smallest singular value
$\sigma_2$ in the main grid as initial condition, but now integrating the
right-singular vector field $\eta_1$ forward and backward in time.

Let us note that the vectorfields $\eta_1, \eta_2$ to be integrated in the
computation of strain and stretch lines are known only in the discrete main
grid. Because of this, the use of a variable step integrator requires
interpolation of the fields and in fact it does not result in better accuracy for the
computed solution. Accordingly, we have selected an order 4 Runge-Kutta (RK4)
integrator with a fixed step, taken smaller than the main grid step.
Moreover the choice of a fixed step RK4 simple integrator not only speeds up
the computations but also handles an added difficulty of the discrete vector
fields $\eta_1, \eta_2$. Pointwise, the vectors $\eta_1(\mx_i), \eta_2(\mx_i)$
are defined up to the sign, which means that the orientation of the field can
suddenly reverse. In practice this is indeed the case, since the SVD algorithm
produces singular vectors that do not vary continuously, but suddenly change
orientation when crossing certain boundaries in the domain. This is unavoidable
when the vector field is not parallelizable~\citep[see][]{Abraham}, which happens
for instance when the two singular values $\sigma_1,\sigma_2$ become equal at a
point in the domain $\Gamma$. We avoid this discontinuity in the sign of the
vectors, which would make the integrator to oscillate back and forth, by asking
our integrator to compare the orientation of the current vector field value
with the previous one used in the integration, and to reverse orientation
of the current vector in case that an orientation discontinuity be detected.

Finally, in order to avoid local extrem values of the singular values
$\sigma_1, \sigma_2$ introduced by fluctuation errors in their computation, our
algorithm fixes a minimal radius such that, only points that are extremal for
the singular value within this radius  are considered as a starting
point for a LCS. Moreover, if one of such points lies within this critical 
distance of an already computed LCS of its type (strainlines for maxima of the
main value $\sigma_1$, stretchlines for minima of the smaller value
$\sigma_2$), then it is discarded as a starting condition of a LCS. The reason
for this is that it would produce a line that is superfluous, since it would be
closely parallel to an already computed LCS of the same type.

\section{LCS in the precessing bar galactic model} \label{sec4:LCSprec}
\sectionmark{LCS in the precessing model}

The purpose of this paper is the computation of Lagrangian Coherent Structures
in both the autonomous precessing bar galactic model and a non-autonomous version,
and to analize and compare the
results with the invariant manifold structure of the same model. Let us start
first with a summary of galactic models and the bar precessing one in
particular.

Barred galaxies represent about $65\%$ of disc galaxies \citep{Eskridge},
characterized mainly by a disc and a central bar-like structure. These
components are usually mathematically modelled by analytical potentials
\citep[][among others]{Athan1983,Pfenn,Pats03,Skokos}. In our work, as in many others,
the disc component is described by a Miyamoto-Nagai potential \citep{Miyamoto},
\begin{equation}
 \phi_d=-\frac{GM_d}{\sqrt{R^2+(A+\sqrt{B^2+z^2})^2}},
 \label{eqn:Miyamotodisc}
\end{equation}
where $R^2=x^2+y^2$ is the cylindrical coordinate radius in the disc plane, and
$z$ denotes the distance in the out-of-plane component. The parameter G is the
gravitational constant and $M_d$ is the mass of the disc while parameters A and
B characterize the shape of the disc. 

The bar structure is modelled by a Ferrers ellipsoid \citep{Ferrers} with
density function,
\begin{equation}
 \rho = 
 \left\lbrace
 \begin{array}{l}
  \rho_0(1-m^2)^{n_h}, \hspace{1cm} m\leq 1 \, , \\
  0,  \hspace{2.8cm} m> 1 \, . \\	
 \end{array}
 \right.
\end{equation}
Here $m^2=x^2/a^2 + y^2/b^2 + z^2/c^2$, and the parameters \emph{a} (semi-major
axis), \emph{b} (intermediate axis), \emph{c}  (semi-minor axis) determine the
shape of the bar while the parameter $n_h$ determines the homogeneity degree for
the mass distribution. The parameter $\rho_0$ is the density at the origin and,
thus, the density distribution $\rho$ depends on the degree of homogeneity $n_h$ 
and the value $\rho_0$ selected.


The precessing bar galactic model \citep{Warps} is a generalization of the
classical galactic model \citep[e.g.:][]{Pfenn,Skokos,Romero1} formed by a
Ferrers bar and a Miyamoto-Nagai disc. 
In the precessing model the bar is assumed to be tilted and precessing in a small 
angle $\varepsilon$ with respect to the galactic plane ($z=0$). So it can be seen as
an order $\varepsilon$ perturbation of the classical model, which is recovered 
when $\varepsilon =0$. A main advantatge of the precessing model is that it
can explain warp structures that appear in many galaxies. 

\begin{figure}
  \centering
    \includegraphics[width=0.49\textwidth]{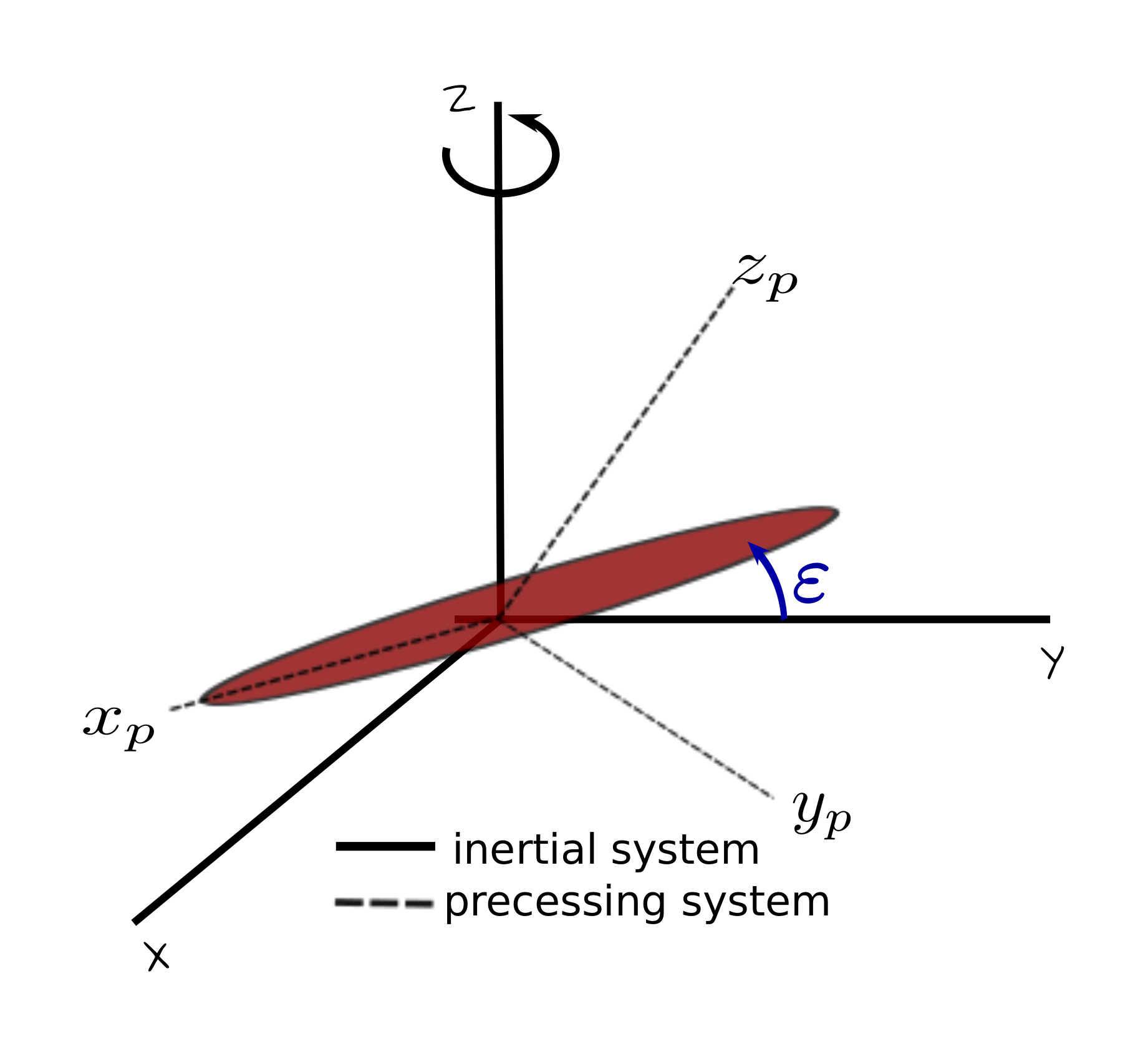}
    \raisebox{1.5cm}{\includegraphics[width=0.49\textwidth]{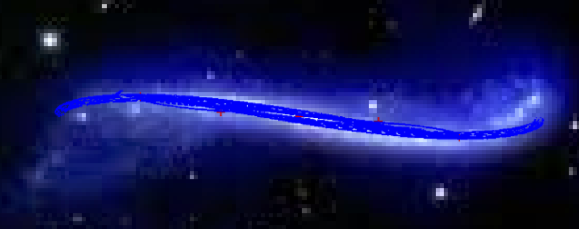}}
  \caption{Top: Precessing model with major axis of the bar aligned with the precessing $x$ axis, and precessing $z$ axis describing a cone about the intertial $Z$ axis. $(x_p,y_p,z_p)$ denotes the axes in the precessing frame and $(X,Y,Z)$ in the inertial one. Bottom: The Integral Sign Galaxy, UGC 3697, with a superposition of the warp obtained from the precessing model.}
  \label{fig:barra_sist_ine}
\end{figure}

In a non-inertial reference frame aligned with the main axis of the ellipsoidal
bar (see Fig.~\ref{fig:barra_sist_ine}) the equations of motion can be written as
the following dynamical system:

\begin{equation} \label{eq:motion}
  \left\lbrace
  \begin{array}{l}
   \ddot{x}=2\Omega \cos(\varepsilon) \dot{y} + \Omega^2 \cos^2(\varepsilon) x + \Omega^2 \sin(\varepsilon)\cos(\varepsilon)z - \phi_{x} \\
   \ddot{y}=-2\Omega \cos(\varepsilon)\dot{x} -2\Omega \sin(\varepsilon)\dot{z} + \Omega^2 y - \phi_{y} \\
   \ddot{z}=2\Omega \sin(\varepsilon)\dot{y} + \Omega^2 \sin(\varepsilon) \cos(\varepsilon) x + \Omega^2 \sin^2(\varepsilon) z - \phi_{z}\\
  \end{array}
 \right.
\end{equation}
where the constant value $\Omega$ is the bar pattern speed and the
potential function $\phi$ is given by the sum of the potentials of the disc and
the bar, $\phi=\phi_d+\phi_b$. We note that when $\varepsilon=0$ we recover the
classical galactic model, and moreover if $\Omega$ is set to one, the dynamics of
the system is someway similar to the so called Restricted Three Body Problem. The detailed derivation of the equations of motion of the precessing model can be found in~\citet{Warps}.

Since our computation of strain and stretch lines is performed on 2-dimensional
domains, while the phase space of the precessing model is 6-dimensional, we are
going to select a simplified, yet physically relevant, case of the model whose
dynamics can be captured by a well placed parametrised surface inside the phase
space $(x,y,z,\dot{x},\dot{y},\dot{z})$.



First, we set the tilt angle $\varepsilon=0$ (classical model), to model an unwarped ringed
galaxy whose dynamics along the $z,\dot{z}$ axes are trivial. Next, we fix the
values $z=0,\dot{z}=0$ to restrict ourselves to the equatorial, or galactic,
plane, which is invariant and captures all the dynamics of the galaxy. The
dynamics of these 2-dimensional galaxy models have been studied in
\citet{Warps,Romero1}. Let us make a brief summary of their results:


\subsection{Dynamics of the precessing model via invariant manifolds}

\begin{figure}
  \centering
    \includegraphics[width=0.31\textwidth]{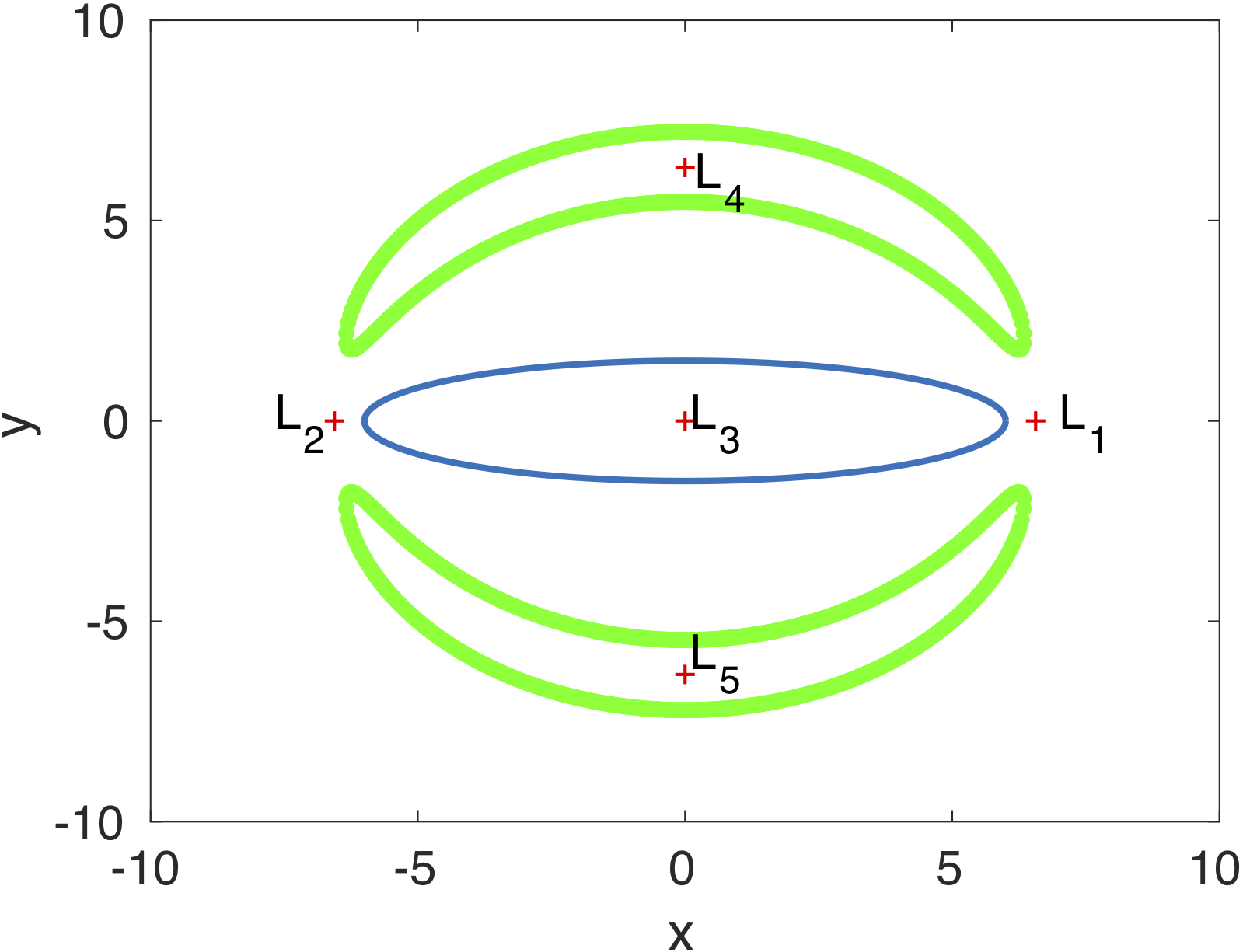}
    \includegraphics[width=0.3\textwidth]{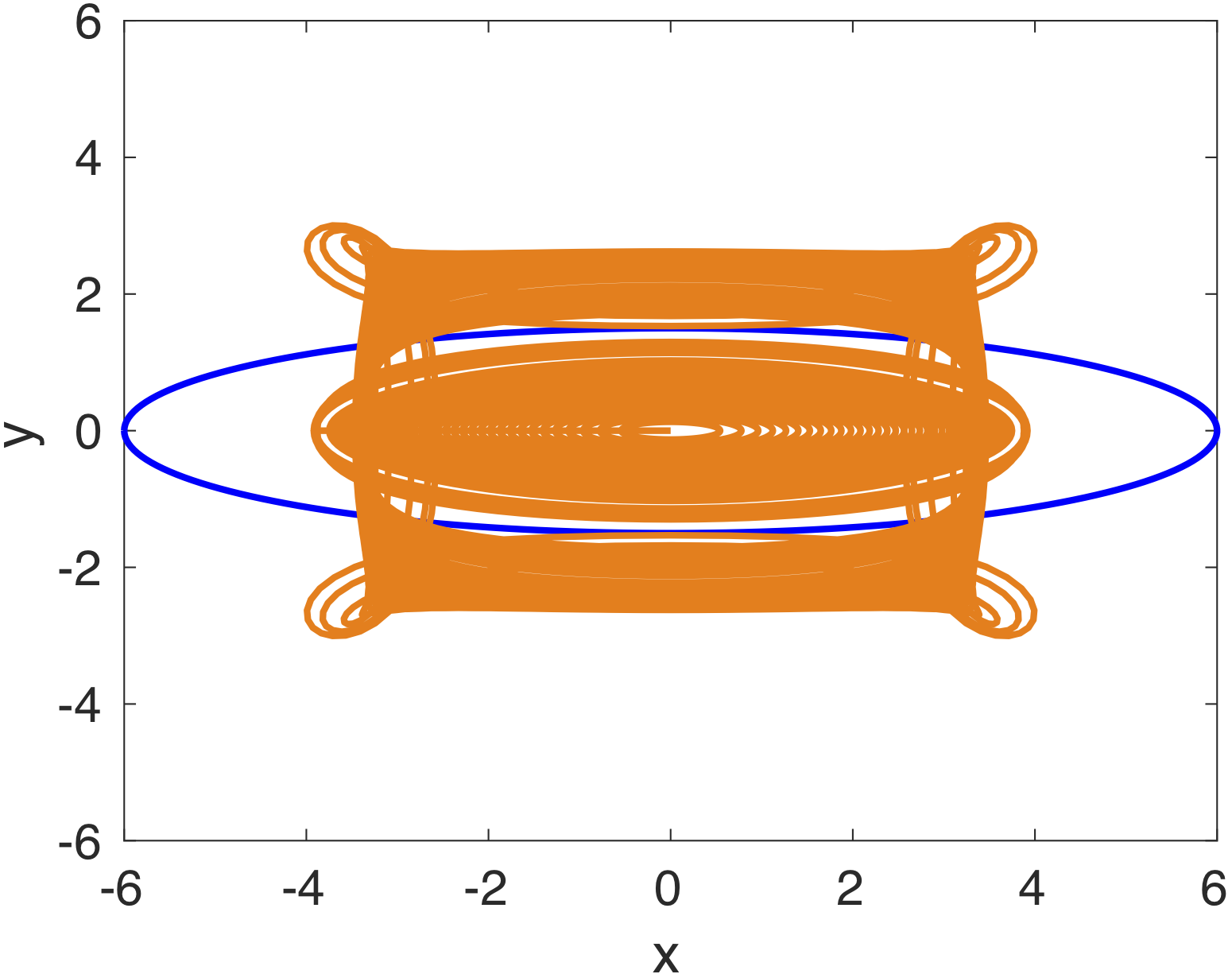}
    \includegraphics[width=0.31\textwidth]{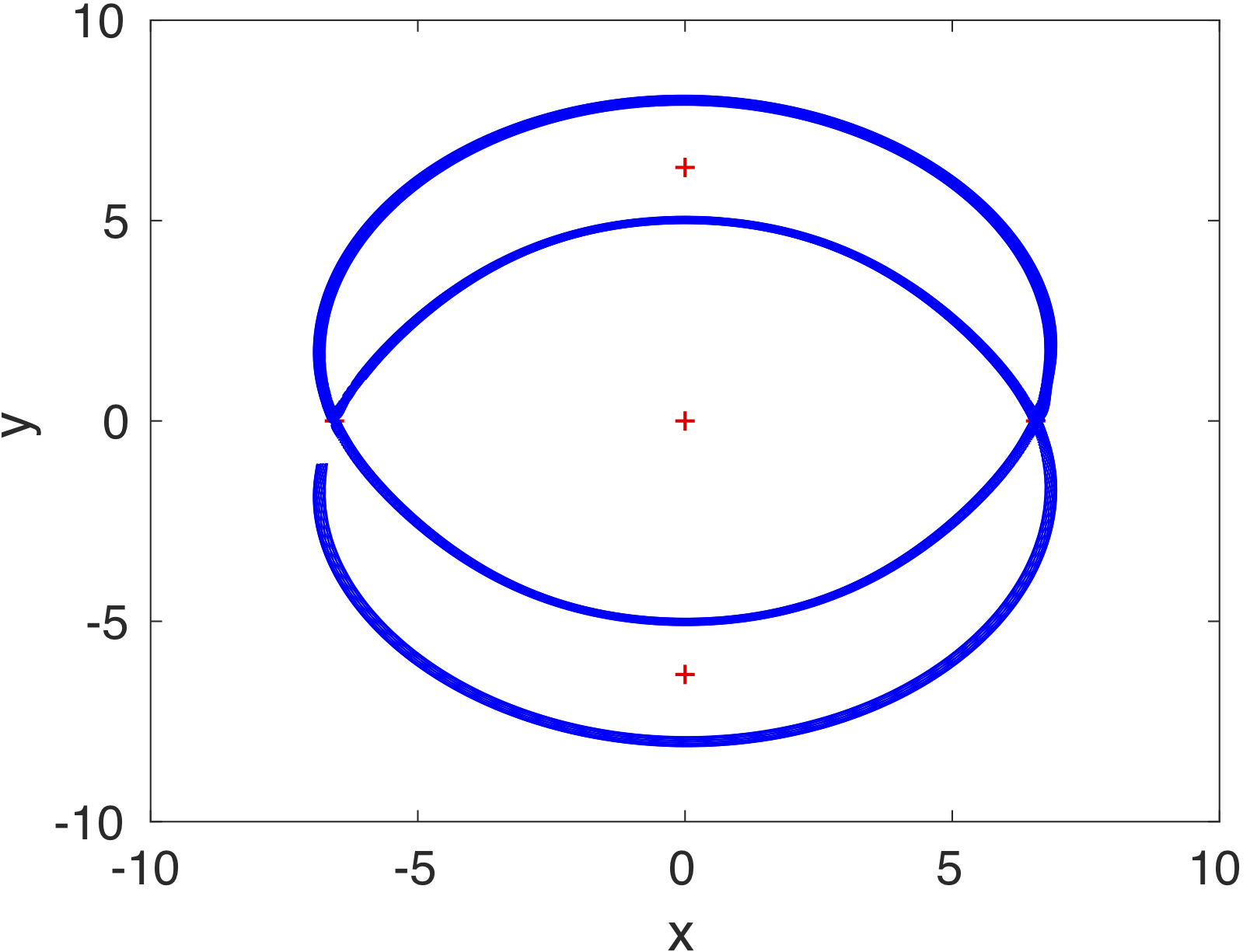}
  \caption{Dynamics in the $xy$ plane of the precessing model with mass bar
$GM_b=0.1$ and tilt angle $\varepsilon=0$, for which the major axis of the bar rotates counterclockwise inside this plane. Top: Lagrange points and zero velocity curves for Jacobi constant $C_J=-0.1876$ (Ferrers bar of the model is
outlined by the blue curve). Center: In orange lines, family of periodic orbits
of the mode, Ferrers bar in blue. Bottom:
In blue, unstable invariant manifolds. In red, Lagrange points of the system.}
\label{fig:peq}
\end{figure}

In this paper we consider the parameters, $A=3, B=1, GM_d=0.9$ for the disc and
$a=6, b=1.5, c=0.6$ for the bar with a pattern speed $\Omega=0.05$, since
these parameters agree with observations and are widely studied \citep{Pfenn}.
The dynamics  are organized around the five equilibrium points ($L_i,
i=1\dots5)$ of the model. In the top panel of Fig.~\ref{fig:peq} we show the
equilibrium points and the zero velocity curves of the system. The points
$L_1$, $L_2$ lie on the x-axis, they are symmetric with respect to the origin,
unstable and surrounded by families of periodic orbits. $L_4$, $L_5$ lie on the y-axis and are surrounded by families of
periodic banana orbits \citep{Athan1983, Contop1981, Skokos}, while $L_3$
lies on the origin of coordinates and is linearly stable. The center panel of
Fig.~\ref{fig:peq} shows the $x_1$ family of planar periodic orbits about $L_3$,
which is mainly stable and has been regarded as responsible for the skeleton of
the bar's structure. But we are particularly interested in the trajectories
outside the bar, driven by the normally hyperbolic invariant manifolds
associated to the libration point orbits about $L_1$ and $L_2$, since they
are responsible of the main visible building blocks in the barred galaxies
through the transport of matter. In the bottom panel of Fig.~\ref{fig:peq} we
can observe these invariant manifolds for the set of parameters taken in the
paper.

Thus, the stable and unstable manifolds of periodic orbits of the same energy
level (from now on denoted by $W_{\gamma_i}^s$ and $W_{\gamma_j}^u$, where
$\gamma_{i}$,$\gamma_{j}$ indicate the periodic orbit), as well as their intersections,
give rise to the responsible structures for the transport of matter in the
galaxy.  In this context, we call heteroclinic orbits the orbits which
correspond to asymptotic trajectories, $\psi$, such that $\psi \in
W_{\gamma_i}^u \cap W_{\gamma_j}^s$, $i \neq j$, $i,j=1,2$, while homoclinic orbits
correspond to the asymptotic trajectories $\psi$ when $i=j$.

The formation of pseudo-rings, rings and spirals in barred galaxies is related,
besides to the invariant manifolds, to the existence of heteroclinic and
homoclinic orbits. Following the classical nomenclature of inner rings (r) and
outer rings (R), outer rings are called, R$_1$ when they have the major axis
perpendicular to the bar major axis, R$_2$ if they have the major axis along
the bar major axis and, R$_1$R$_2$ when they have a component parallel and
another one perpendicular to the bar. As explained in \citet{Romero1,Romero2},
the formation of rR$_1$ rings is linked to the existence of heteroclinic
orbits, R$_1$R$_2$ is linked to the existence of homoclinic ones, while spiral 
arms and R$_2$ rings appear when there exist neither heteroclinic nor
homoclinic orbits.  

\begin{figure}
  \centering
    \includegraphics[width=0.3\textwidth,angle=0]{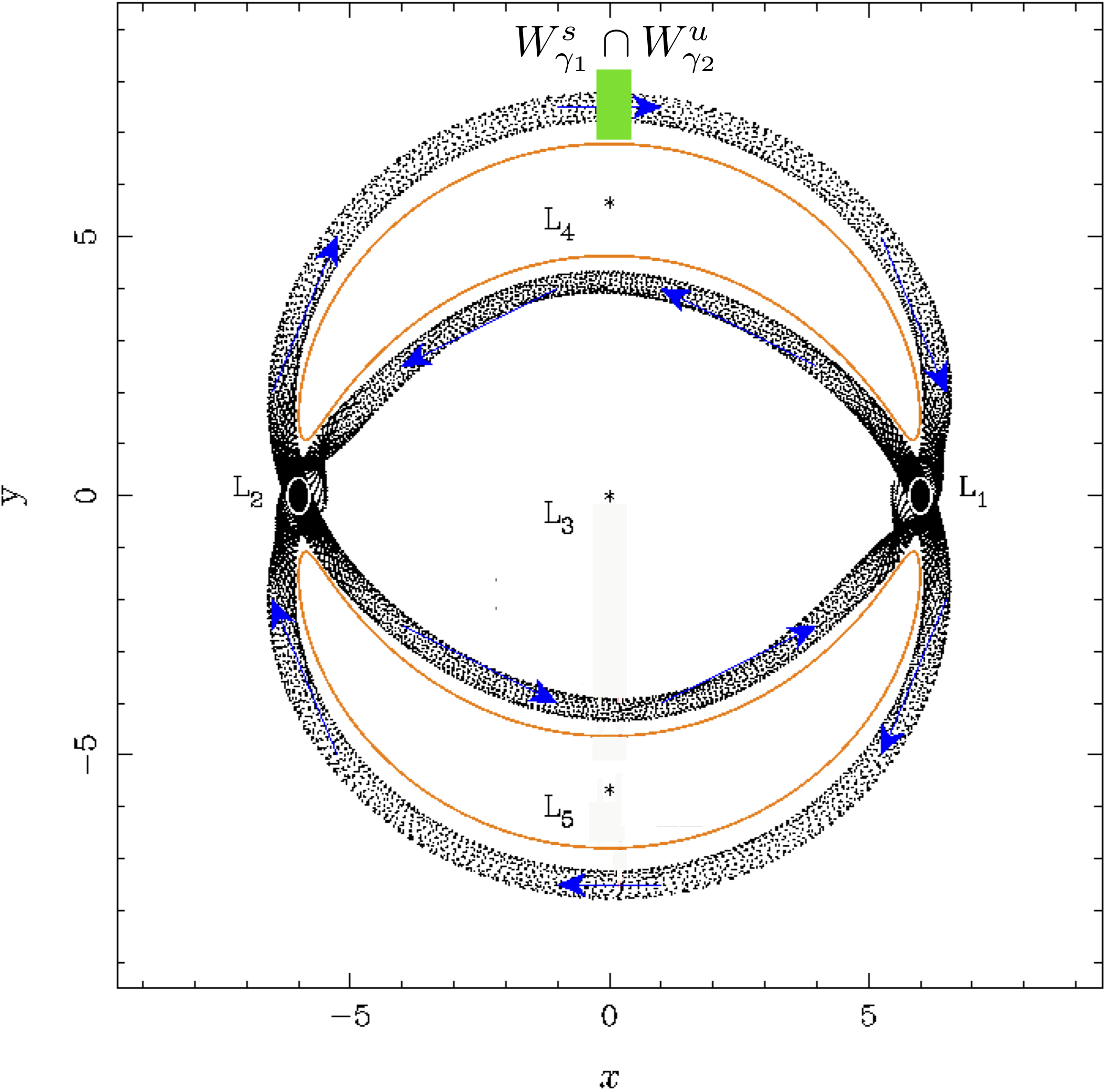}
    \includegraphics[width=0.3\textwidth,angle=0]{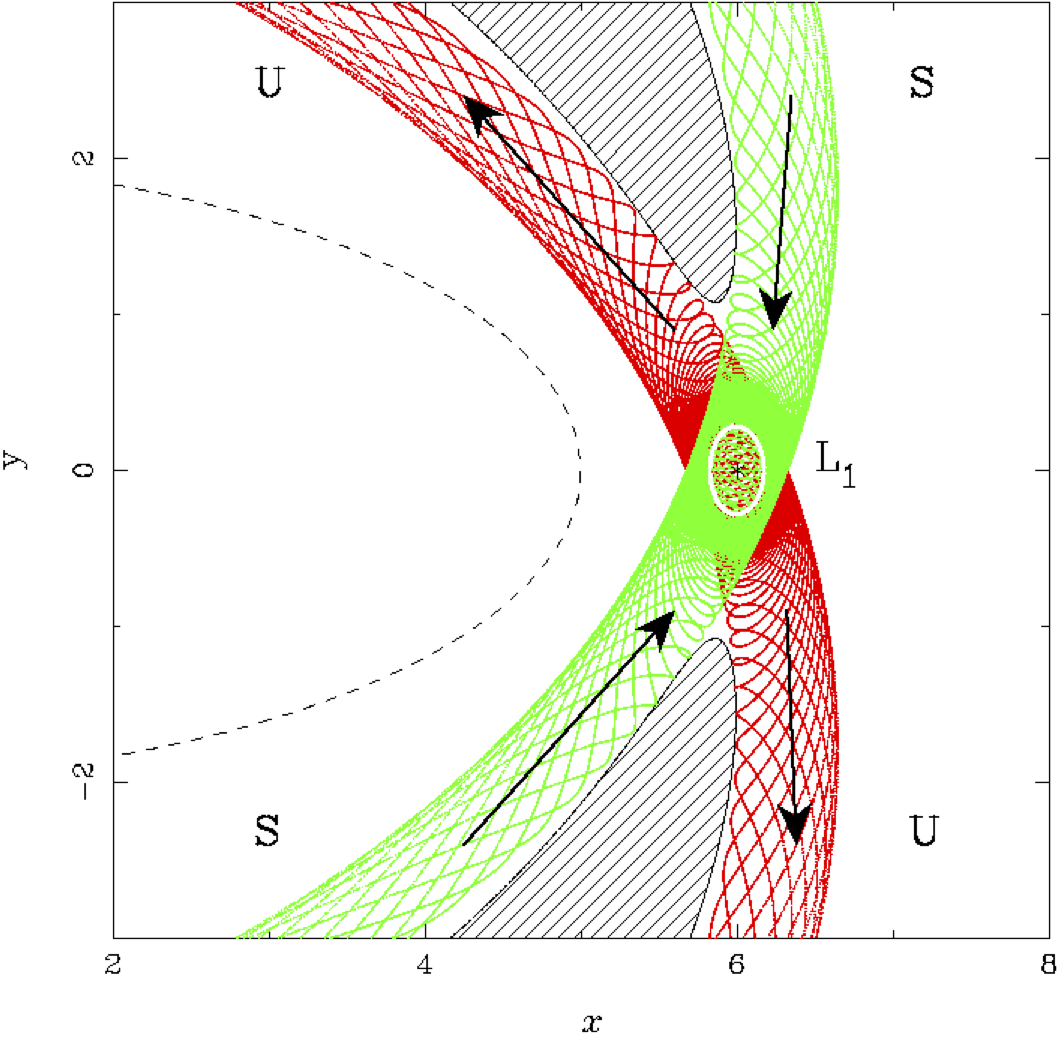}
    \includegraphics[width=0.3\textwidth,angle=0]{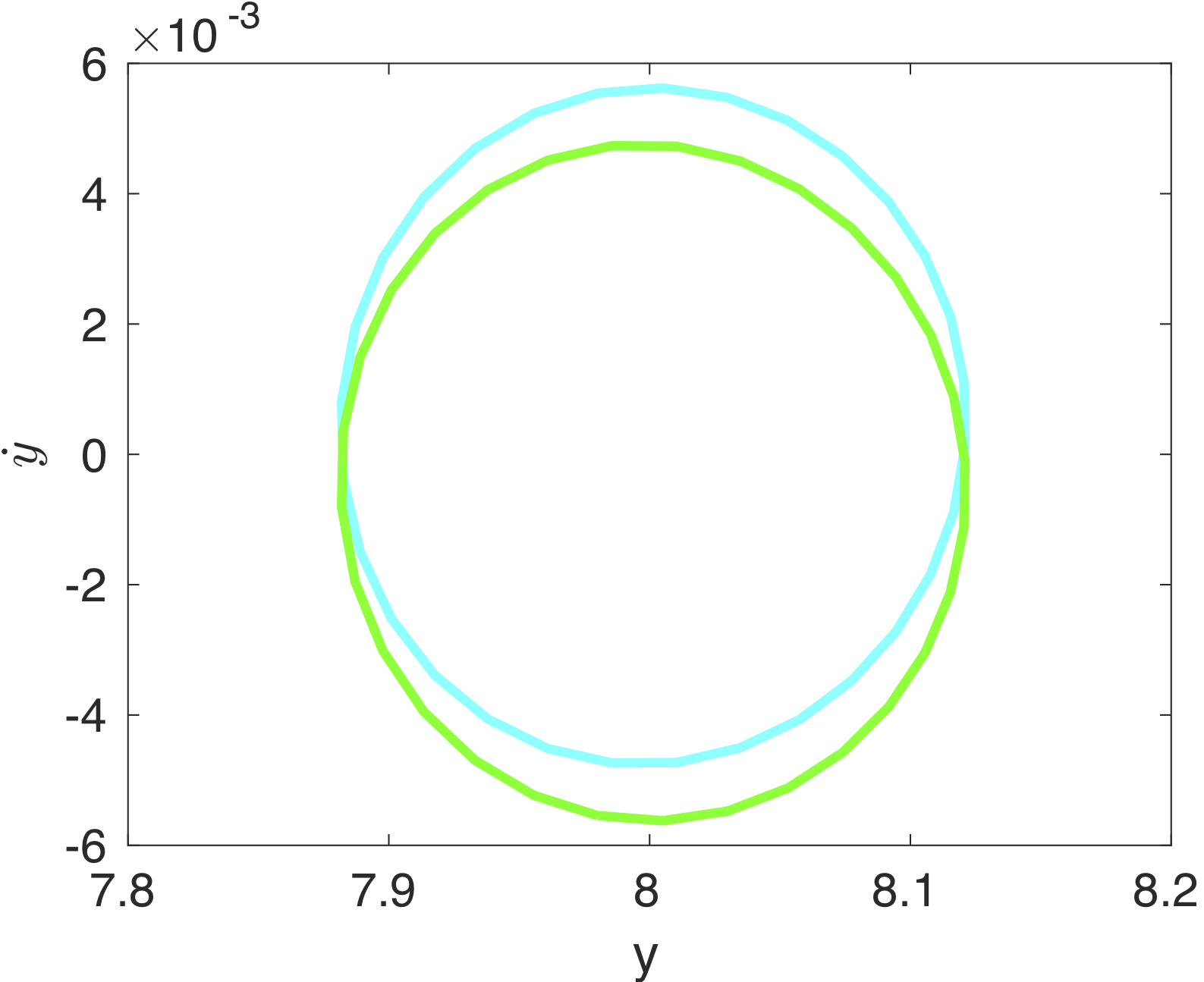}
    \caption{Dynamics of a rR$_1$ ringed galaxy. Top: In blue, arrows
indicating circulation of matter. In green, intersection of the invariant manifolds with hyperplane $S$.
All of them in the $(x,y)$ plane. (Image adapted from~\citet{Romero3}). Center:
Neighborhood of equilibrium point L$_1$ in $(x,y)$ plane. Arrows: Motion along
invariant manifolds (S=stable, U=unstable). Stripped areas: Forbidden regions
delimited by zero velocity curves. (Image taken from~\citet{Romero3}). Bottom:
Precessing model with masses $GM_b=0.1$, $GM_d=0.9$, pattern speed
$\Omega=0.05$, and tilt angle $\varepsilon=0$. We display the $(y,\dot{y})$
projections of the first crossings of the unstable and stable manifolds with the
hyperplane $S$. In cyan: $W_{\gamma_1}^{s,1}$, in green: $W_{\gamma_2}^{u,1}$.
See more details in the text.} 
\label{fig:recorrido_heteroc}
\end{figure}

So for instance, the existence of these heteroclinic orbits makes the galaxy a
rR$_1$ ringed galaxy, because they establish a closed path along which the
matter is transported (Fig.~\ref{fig:recorrido_heteroc}).
The transfer of matter happens mainly from the inner region delimited between
the bar and the zero velocity curves to the outer region. The transit orbits
contained inside the manifold tubes (\citet{Romero1,Romero2,GideMasd})
are responsible for this action. On the other hand, the non-transit orbits are
those that stay out of the manifold tube and move only around the bar without
going out to the outer regions.

As the dynamics of our system takes place in a six dimensional phase space, we
compute the intersections of the trajectories of the invariant manifolds with
the hyperplane $S$ given by the section $x=0$ in phase space. We consider the
outer branches of the stable invariant manifold of the Lyapunov orbit around
L$_1$, $W_{\gamma_1}^s$, and the unstable invariant manifold of the Lyapunov
orbit around L$_2$, $W_{\gamma_2}^u$, both at the same energy level. The first
intersection of these two invariant manifolds with the hyperplane $S$ are two
closed curves. Considering the $y\,\dot{y}$ projection, we denote by
$W_{\gamma_1}^{s,1}$ the closed curve resulting from the first intersection of
$W_{\gamma_1}^s$, and by $W_{\gamma_2}^{u,1}$ the closed curve resulting from
the first intersection of $W_{\gamma_2}^u$. The intersection
$W_{\gamma_1}^{s,1} \cap W_{\gamma_2}^{u,1}$ corresponds to heteroclinic orbits
for the given energy level of the invariant manifolds. Analogously, the second
intersection of the invariant manifolds $W_{\gamma_1}^s$, $W_{\gamma_2}^u$ with
the hyperplane $S$ are denoted by $W_{\gamma_1}^{s,2}$ and $W_{\gamma_2}^{u,2}$
respectively.

When the tilt angle $\varepsilon$ takes the value $\varepsilon=0$, the plane
$z=0$ is invariant, so the phase space is reduced to four dimensions, which,
together with the fixed energy level let us define a state just selecting
a $(y,\dot{y})$ point on $S$. (This is, a point on $S$ is defined by $x=0$,
and since $z=\dot{z}=0$, selecting $y$ and $\dot{y}$, for the fixed energy level
and the sense of crossing, also $\dot{x}$ is determined, completing this way the 
state).

Then the points on the curve $W_{\gamma_1}^{s,1}$ correspond to states
on $W_{\gamma_1}^{s}$ and so they are orbits that tend asymptotically to the
Lyapunov orbit $\gamma_1$.  In the same way, the points on the curve
$W_{\gamma_2}^{u,1}$ provide states on $W_{\gamma_2}^{u}$, therefore they are
orbits that depart asymptotically from the Lyapunov orbit $\gamma_2$. This
means that when $\gamma_1$ and $\gamma_2$ are both in the same energy level,
the intersection points $W_{\gamma_1}^{s,1} \cap W_{\gamma_2}^{u,1}$ correspond
to heteroclinic orbits between them.  In the bottom panel of
Fig.~\ref{fig:recorrido_heteroc} we observe that for $\varepsilon=0$ there are
two heteroclinic orbits corresponding to the intersection of
$W_{\gamma_1}^{s,1}$ and $W_{\gamma_2}^{u,1}$. 

As it is known, the $(y,\dot{y})$ points outside both curves,
$W_{\gamma_1}^{s,1}$ and $W_{\gamma_2}^{u,1}$, correspond to states whose
trajectories remain inside the inner region of the galaxy delimited by the zero
velocity curves, i.e. they are non-transit orbits. Finally the $(y,\dot{y})$
points that are inside the intersection defined by both curves correspond to
orbits that transit from the inner region to the outer one, i.e. they are
transit orbits. It is in this way how the invariant manifolds of the Lyapunov
orbits drive the motion of the stars from the inner to the outer regions. See
Fig.~\ref{fig:recorrido_heteroc} and also \citet{tesis} for many more details.

\subsection{Dynamics of the precessing model via Lagrangian Coherent Structures}

In order to obtain a surface containing the heteroclinic orbits between the
L$_1$ and L$_2$ regions of our model, we fix an energy level $C_{L_1}+\delta$
slightly above of that of the equilibrium point L$_1$ (or L$_2$) and we set $x=0$.
As we have already mentioned, the states of the system on this surface can be
parameterised by $(y,\dot{y})$ since using the Jacobi constant of the system,
 \begin{equation}
 \begin{split}
    C(x,y,z,\dot{x},\dot{y},\dot{z}) =& -(\dot{x}^2+\dot{y}^2+\dot{z}^2)+2\Omega^2\sin(\varepsilon)\cos(\varepsilon)xz \\
   & +\Omega^2(\cos^2(\varepsilon)x^2+y^2+\sin^2(\varepsilon)z^2)-2\phi,
 \end{split} 
 \label{eqn:jacobicte}
\end{equation}
and taking into account that for planar orbits $z=\dot{z}=0$, we obtain 
$\dot{x}$ as,
\begin{equation}
 \dot{x} = \sqrt{-\dot{y}^2+\Omega^2 y^2-2\phi(0,y,0)-(C_{L_1}+\delta)}.
\end{equation}
In this way we have a surface $S \subset \mathbb{R}^6$ parametrised by 
\begin{equation}
\Psi(y,\dot{y}) = (0,y,0,\sqrt{-\dot{y}^2+\Omega^2 y^2-2\phi(0,y,0)-(C_{L_1}+\delta)},\dot{y},0),
\end{equation}
where we have chosen the positive value of $\dot{x}$, to obtain the plane
containing the above mentioned heteroclinic orbits (from the L$_2$ neigbourhood
to the L$_1$ one, this is, we cross $S$ from $x<0$ to $x>0$).

\begin{figure}
    \includegraphics[width=0.45\textwidth]{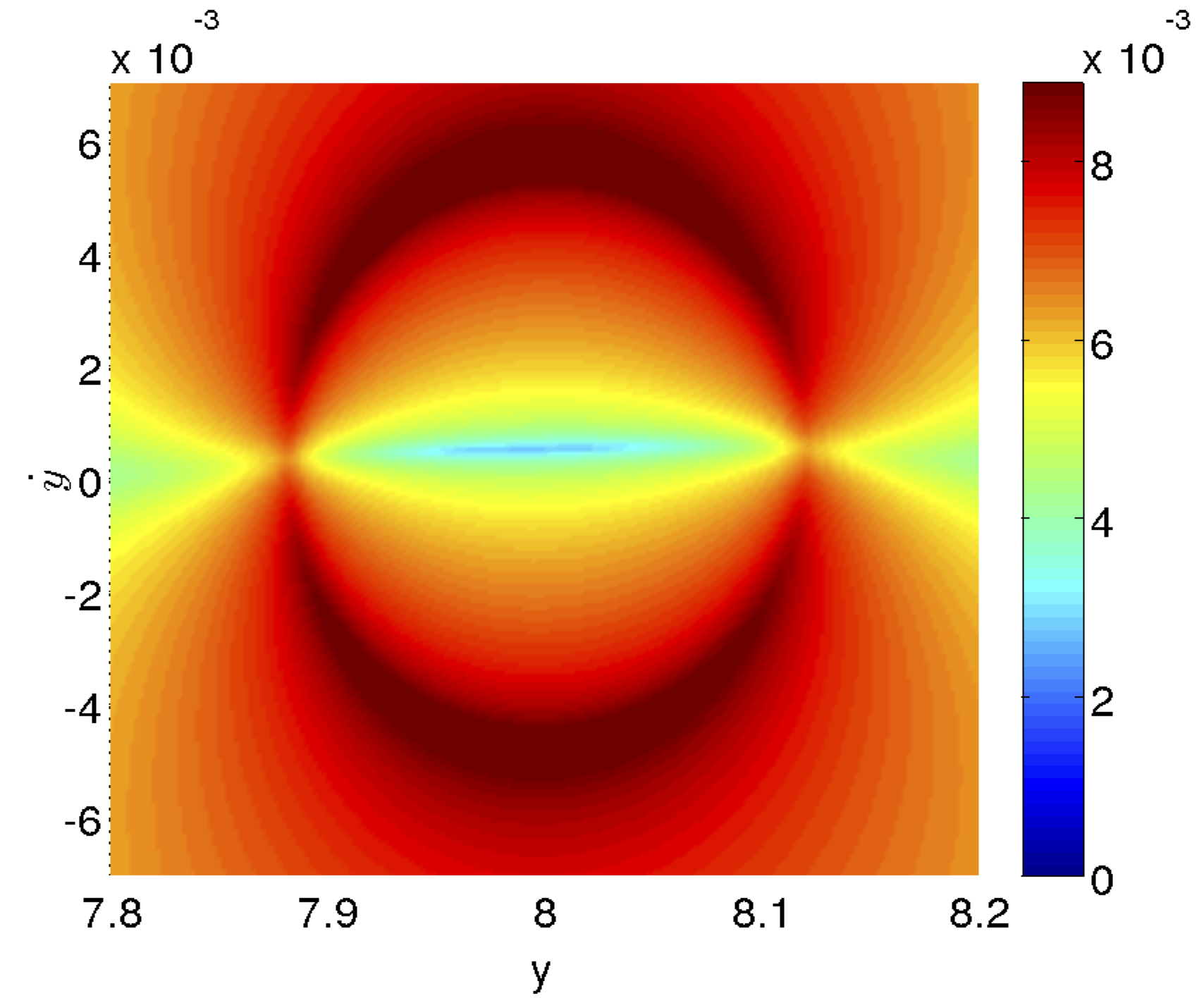} 
    \includegraphics[width=0.45\textwidth]{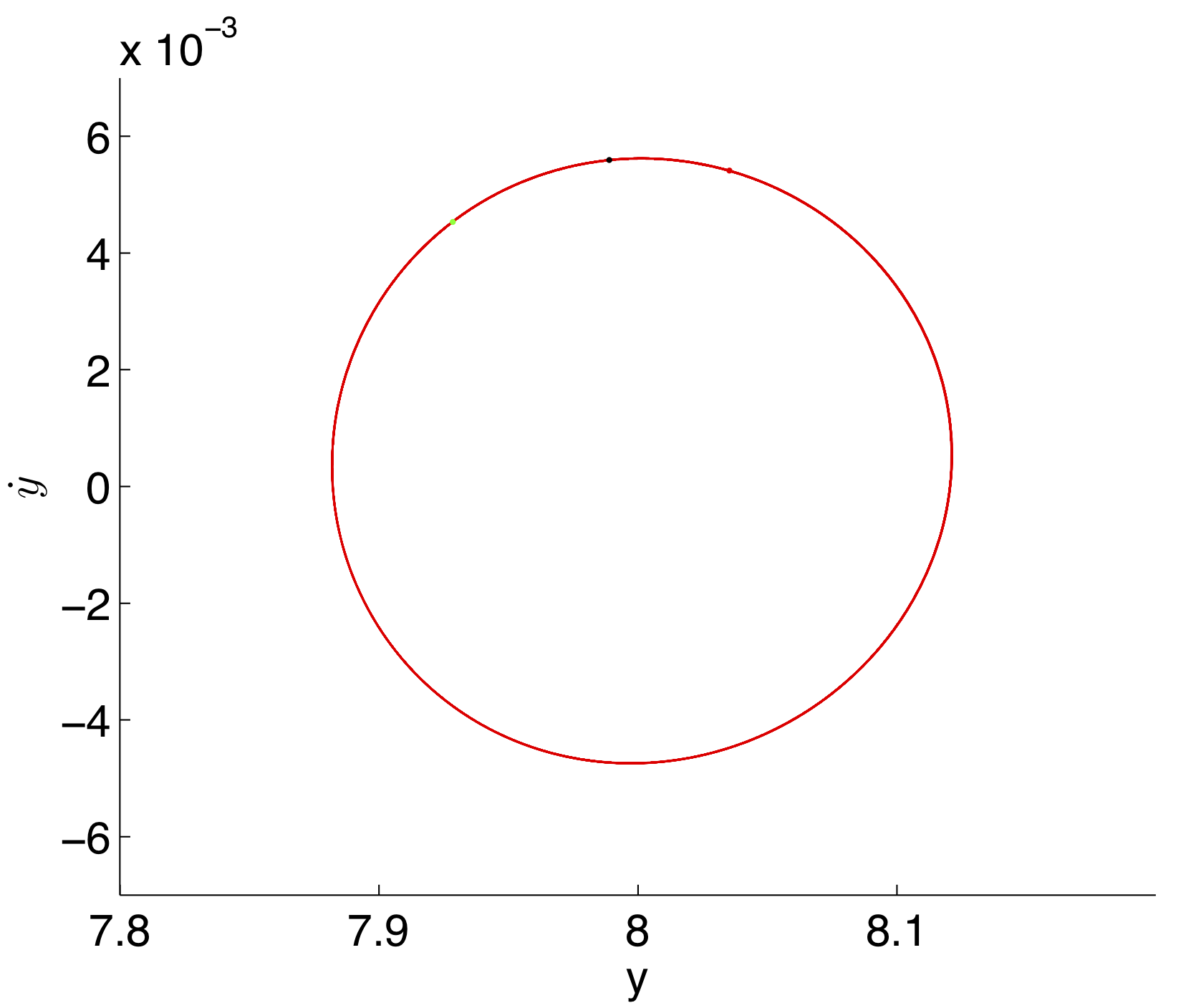} 
    \caption{FTLE (top) and strainlines (bottom) of the classical model for time $T=[0,505]$.}
  \label{fig:strainprec505}
\end{figure}

We consider the same parameters as in the previous case. Unless otherwise
indicated, the spatial domain for our parametrization is $\Gamma =
[7.8,8.2]\times[-0.007,0.007]$. Fig~\ref{fig:strainprec505} shows the FTLE
field and the strainlines corresponding to this model for an integration time
$T=[0,505]$, where $t=505$ is the time of the first intersection of the upper
branch of the stable invariant manifold with the $x=0$ plane. Note that
the concentration of strainlines is found on the main ridges of the FTLE field,
forming a closed curve.  

\begin{figure}
\centering
    \includegraphics[width=0.32\textwidth]{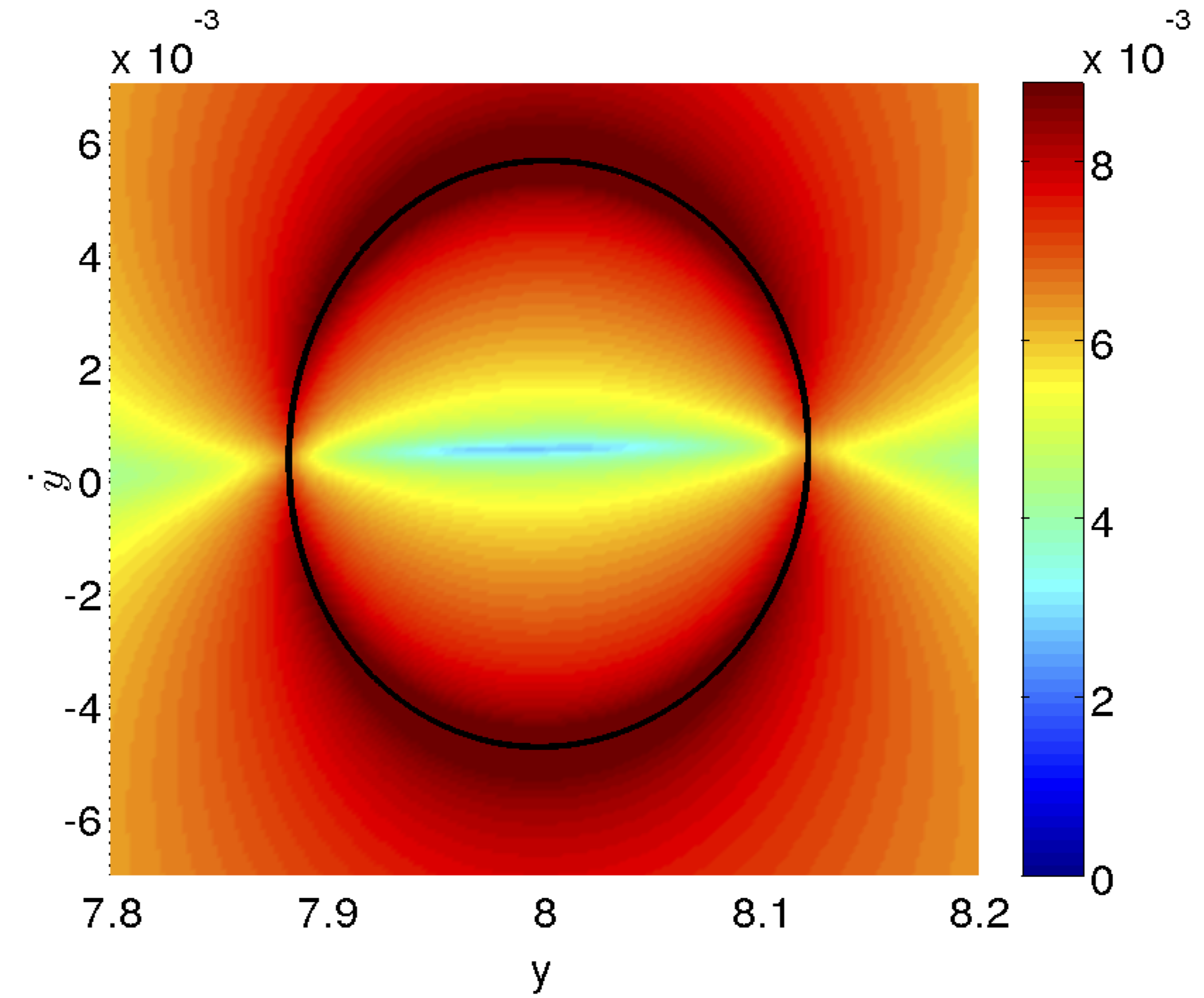} 
    \includegraphics[width=0.34\textwidth]{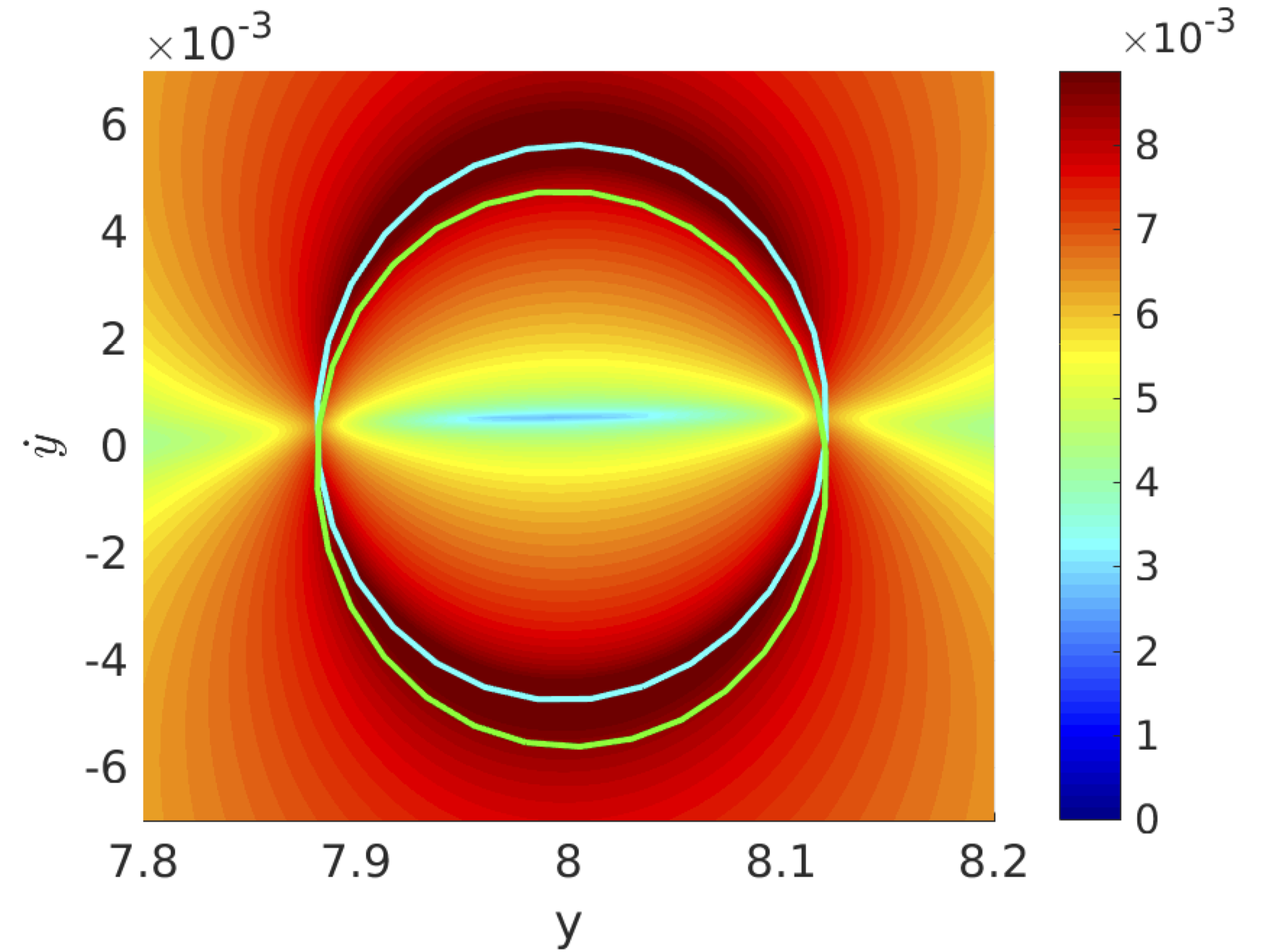} 
    \includegraphics[width=0.34\textwidth]{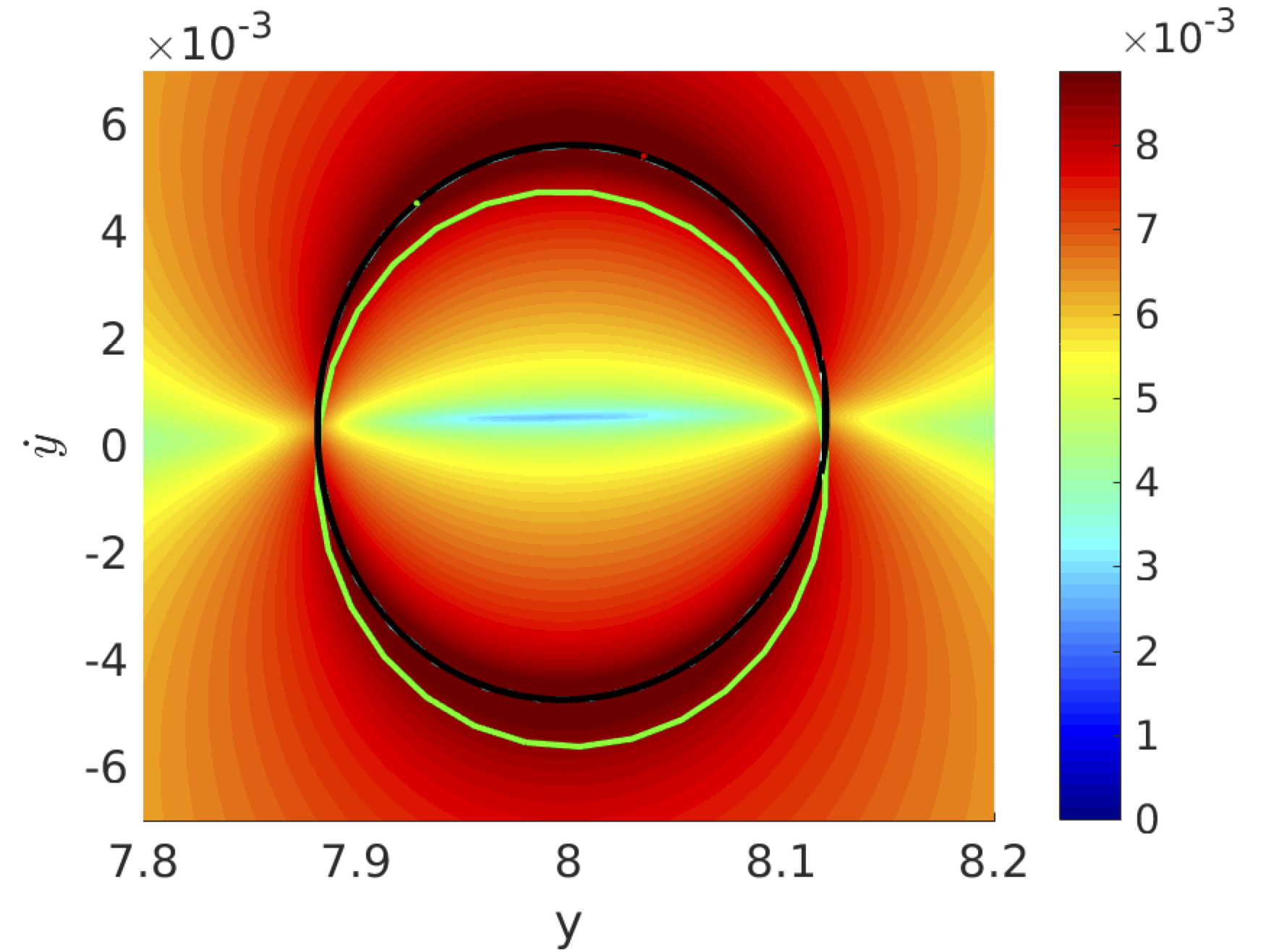} 
    \caption{Classical model for time
$T=[0,505]$. Top: FTLE field and strainlines in black. Center: FTLE
field and heteroclinic orbits ($W_{\gamma_1}^{s,1}$ in cyan,
$W_{\gamma_2}^{u,1}$ in green). Bottom: FTLE field, strainlines (in black) and
heteroclinic orbits (in cyan and green). Cyan and black coincide perfectly.}
\label{fig:strainheterprec505}
\end{figure}

Let us overimpose the FTLE field, the strainlines and the first cuts of the
heteroclinic orbits with the $\{x=0,\, y\ge 0\}$ semiplane, for the same level
of energy ($C_{L_1}+\delta$, with $\delta$ small).
Figure~\ref{fig:strainheterprec505} shows in the top panel how the
strainlines follow the main ridge of the FTLE field. If we observe
$W_{\gamma_1}^{s,1}$ and $W_{\gamma_2}^{u,1}$ (center panel), the one
corresponding to the stable manifold follows as well the main ridge of the FTLE
field, whereas the one corresponding to the unstable manifold is associated
with the boundary of the ridge. When we join both figures, in the bottom panel,
the strainlines correspond exactly with the heteroclinic orbit of the stable
manifold. Therefore, we confirm that the strainlines are associated with the
stable invariant manifolds, and more precisely with $W_{\gamma_1}^{s,1}$, at 
least in the autonomous problem. Let us remark that as the stretchlines are 
by definition perpendicular to the strainlines, they do not seem to carry any
relevant dynamic information in our problem.

\begin{figure}
    \includegraphics[width=0.45\textwidth]{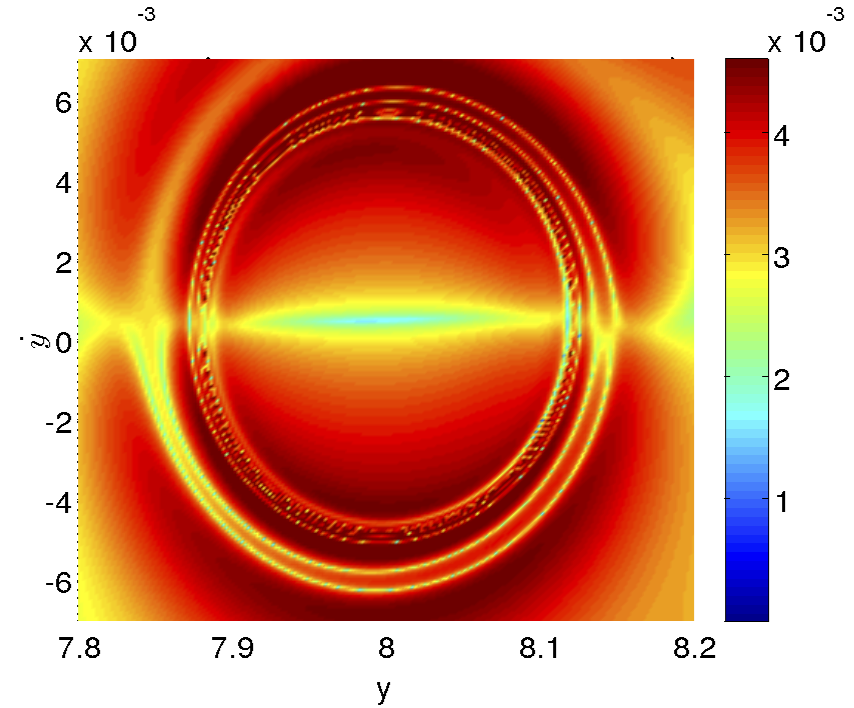} 
    \includegraphics[width=0.45\textwidth]{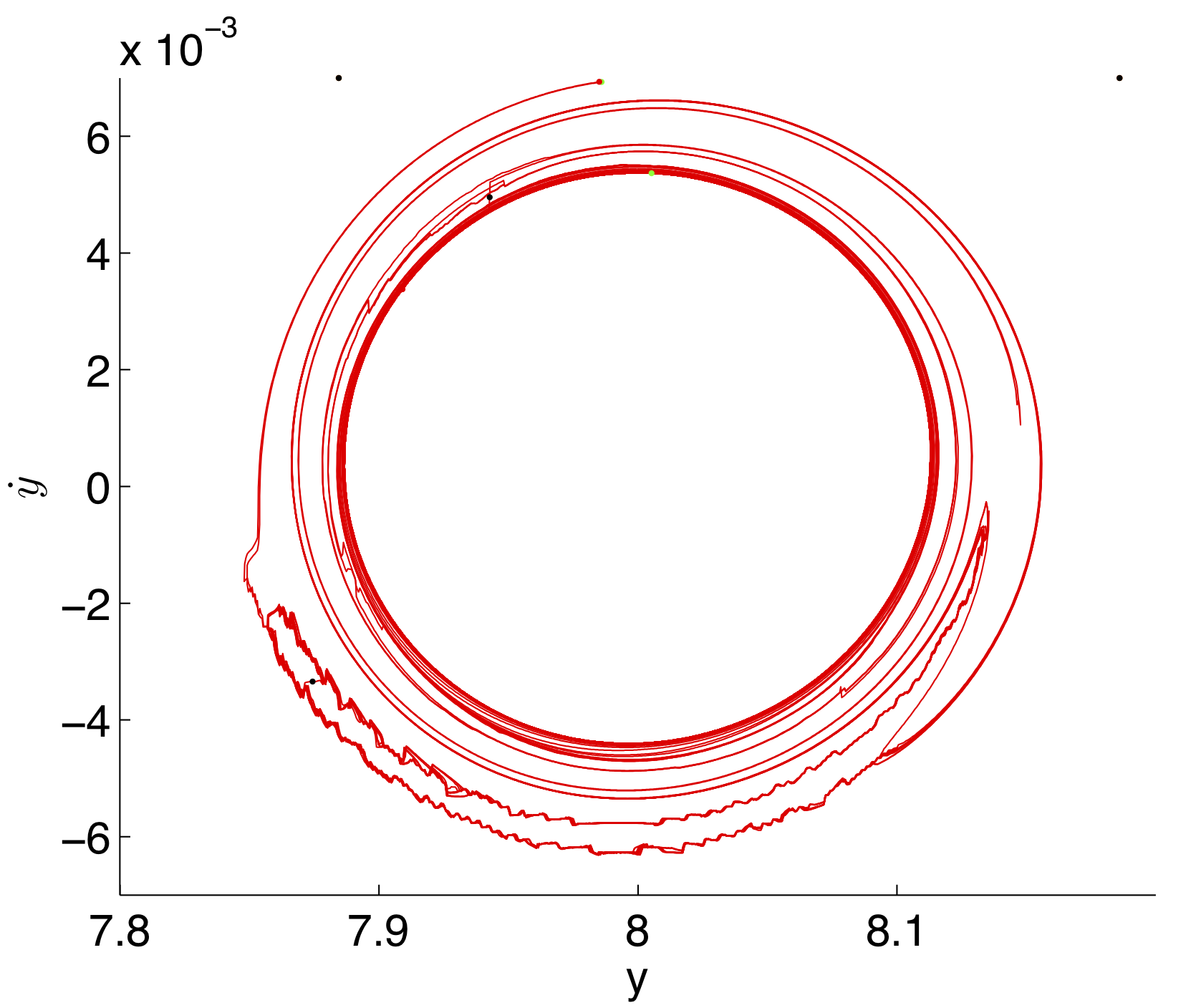} 
    \caption{FTLE (top) and strainlines (bottom) of the classical model for time $T=[0,1000]$.}
  \label{fig:strainprec1000}
\end{figure}

When the integration time is increased, the accuracy of the description of the
FTLE field and strainlines is gradually lost. This is due not only to the loss
of accuracy in the integration of the dynamical system when computing the
LCS, but also to the fact that, in the classical model, the successive
intersections of the invariant manifolds with the semiplane $\{x=0,\, y\ge 0\}$ 
become increasingly blurred due to the transversal intersection of manifolds
and the inherent chaotic dynamics (see \citet{GideMasd}). Taking an integration
time $T=[0,1000]$ the ridges of the FTLE field are not so remarked, but the
strainlines continue following these ridges, although the dynamics of the
system is less clear (see Fig.~\ref{fig:strainprec1000}).

\begin{figure}
    \includegraphics[width=0.45\textwidth]{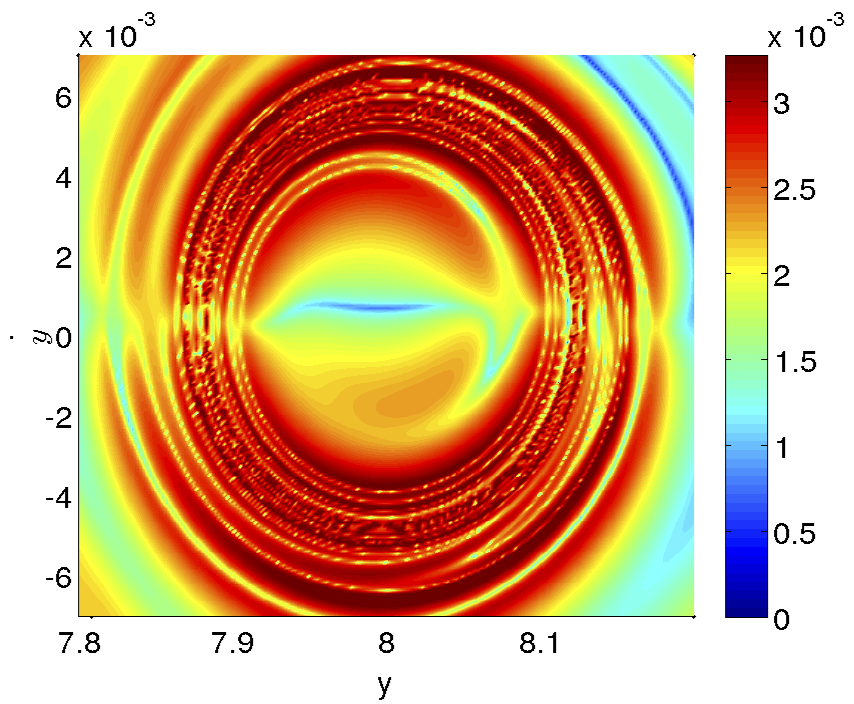} 
    \includegraphics[width=0.45\textwidth]{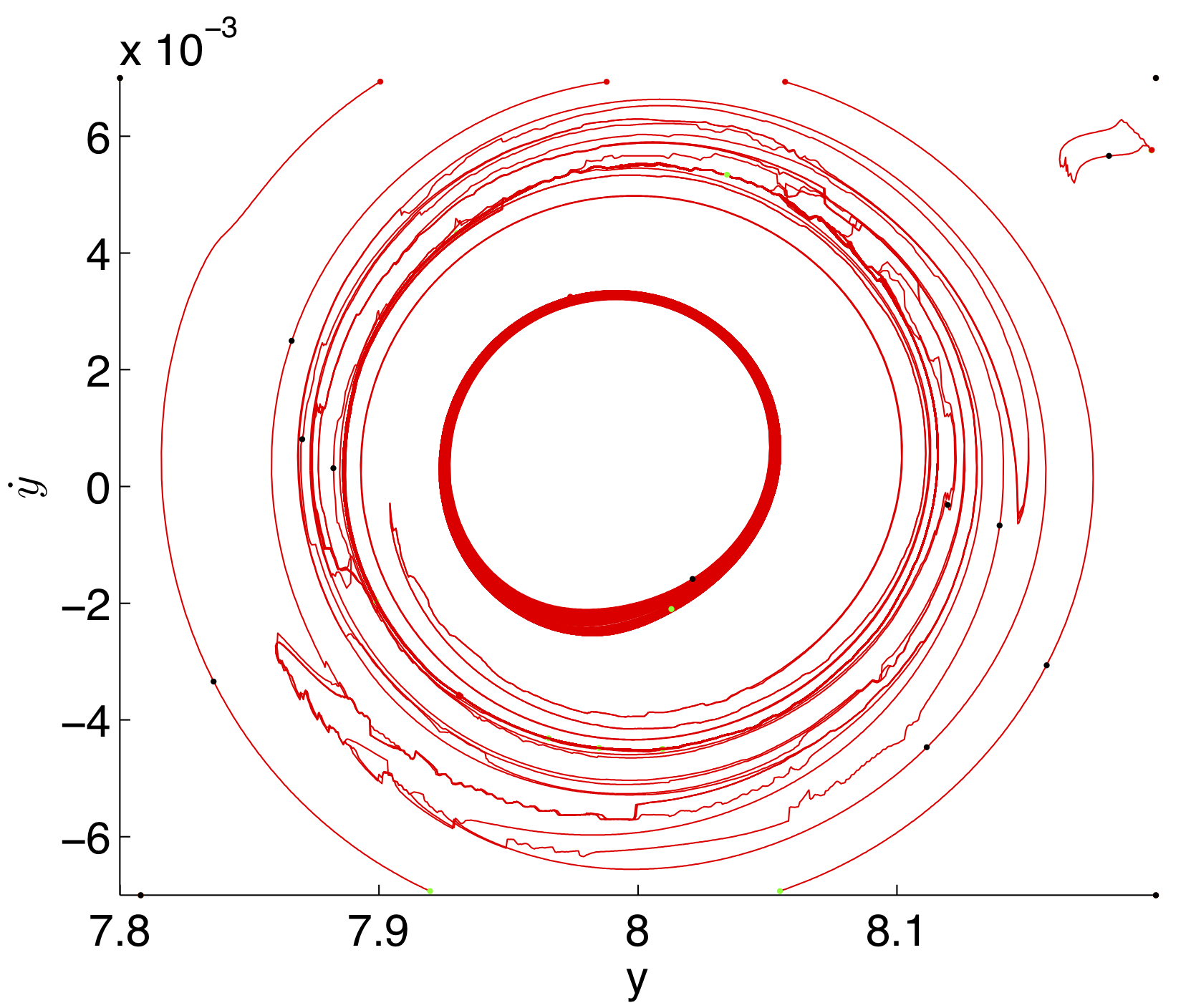} 
    \caption{FTLE (top) and strainlines (bottom) of the classical model for time $T=[0,1570]$.}
  \label{fig:strainprec1570}
\end{figure}

In order to observe the second intersection of the invariant manifolds with the
semiplane $\{x=0,\, y\ge 0\}$, $W_{\gamma_1}^{s,2}$ and $W_{\gamma_2}^{u,2}$,
and the corresponding LCS, we take a time interval for the integration of
$T=[0,1570]$. Figure~\ref{fig:strainprec1570} shows the FTLE field, where the
ridges are less marked than in the previous case, as well as the strainlines
associated with these ridges, but still the impact of the chaotic dynamics
is noticed with some structure. However we also observe some artifacts in the
integration of strainlines that do not correspond to ridges of the FTLE field,
but are due to the long time of integration, for example in the top right part
of the figure of strainlines.

\begin{figure}
\centering
    \includegraphics[width=0.32\textwidth]{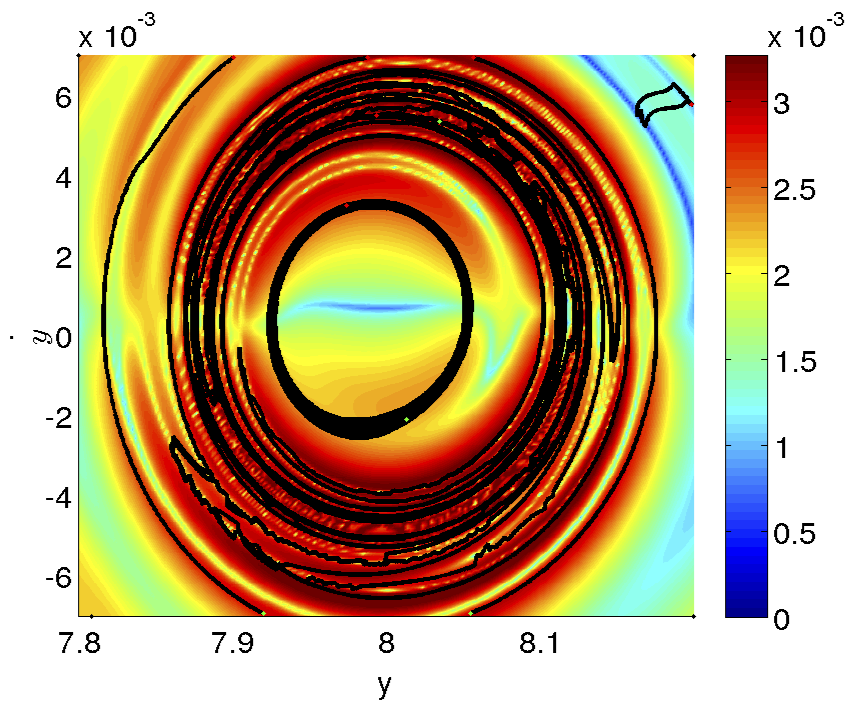} 
    \includegraphics[width=0.32\textwidth]{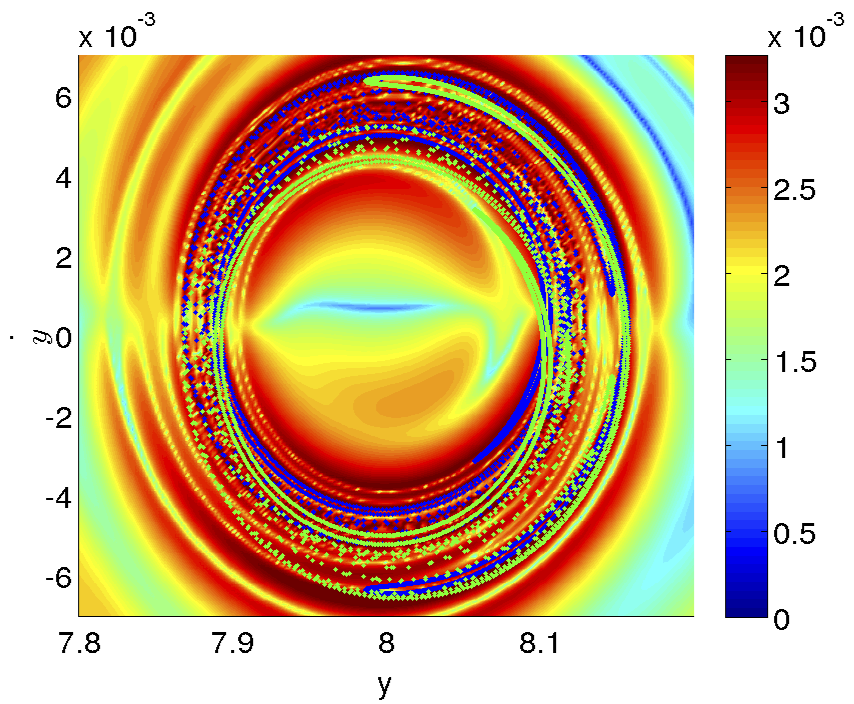} 
    \includegraphics[width=0.32\textwidth]{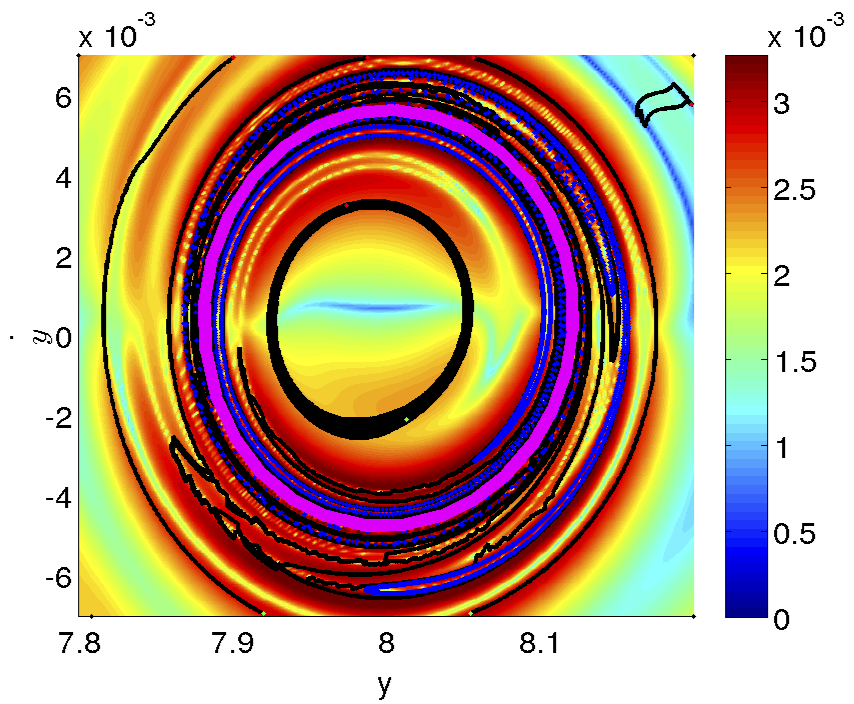} 
    \caption{Classical model for time
$T=[0,1570]$. Top: FTLE field and strainlines in black. Center: FTLE
field and heteroclinic orbits for the second intersection of the $\{x=0,\, y\ge
0\}$ semiplane ($W_{\gamma_1}^{s,2}$ in blue, $W_{\gamma_2}^{u,2}$ in green).
Bottom: FTLE field, strainlines (in black) and heteroclinic orbits for the
first intersection (in magenta), $W_{\gamma_1}^{s,1}$, and for the second one
(in blue), $W_{\gamma_1}^{s,2}$, of the stable manifold.}
\label{fig:strainheterprec1570}
\end{figure}

In Fig.~\ref{fig:strainheterprec1570} we observe how the strainlines (in black)
follow the main ridges of the FTLE field, although some of the strainlines do
not match these ridges (top panel). In the center panel we show the
cuts of the heteroclinic orbits, $W_{\gamma_1}^{s,2}$ and $W_{\gamma_2}^{u,2}$,
superimposed to the FTLE field, where we mark this second intersection of the
invariant manifolds with dots in order to clarify the figure, since they do not
follow a clear closed curve. We observe that the second intersection of the
stable manifold (in blue) follows approximately the ridges of the FTLE field.
To join this stable manifold to the strainlines and the FTLE field (bottom
panel), the stable manifold (the first intersection with the plane $x=0$ in
magenta, the second one in blue) is closely approximated by the strainlines and
the ridges of the FTLE field in its main components. Let us point out that both
the strainlines and the ridges of the FTLE field also give false positives as
time increases, i.e. not all the strainlines follow the ridges of the FTLE
field and furthermore some ridges of the FTLE field do not correspond to a
defined structure of the dynamics of the model. This fact suggests that for
long integration times the FTLE field and the strainlines lose its precision, a
fact that is also stated by, e.g.,~\citet[][]{Haller12}. 

\begin{figure}
    \includegraphics[width=0.45\textwidth]{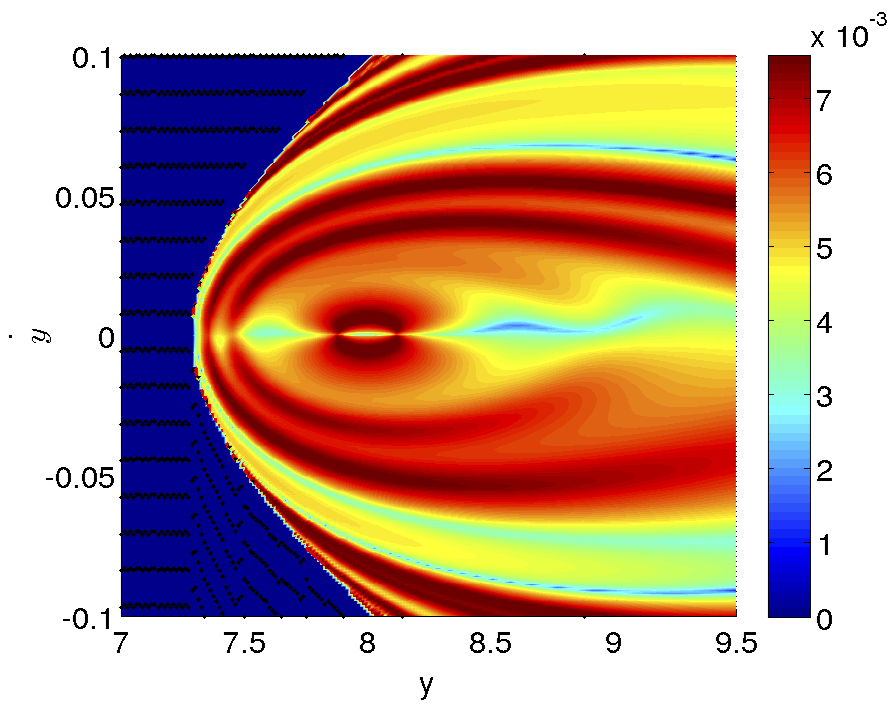} 
    \includegraphics[width=0.45\textwidth]{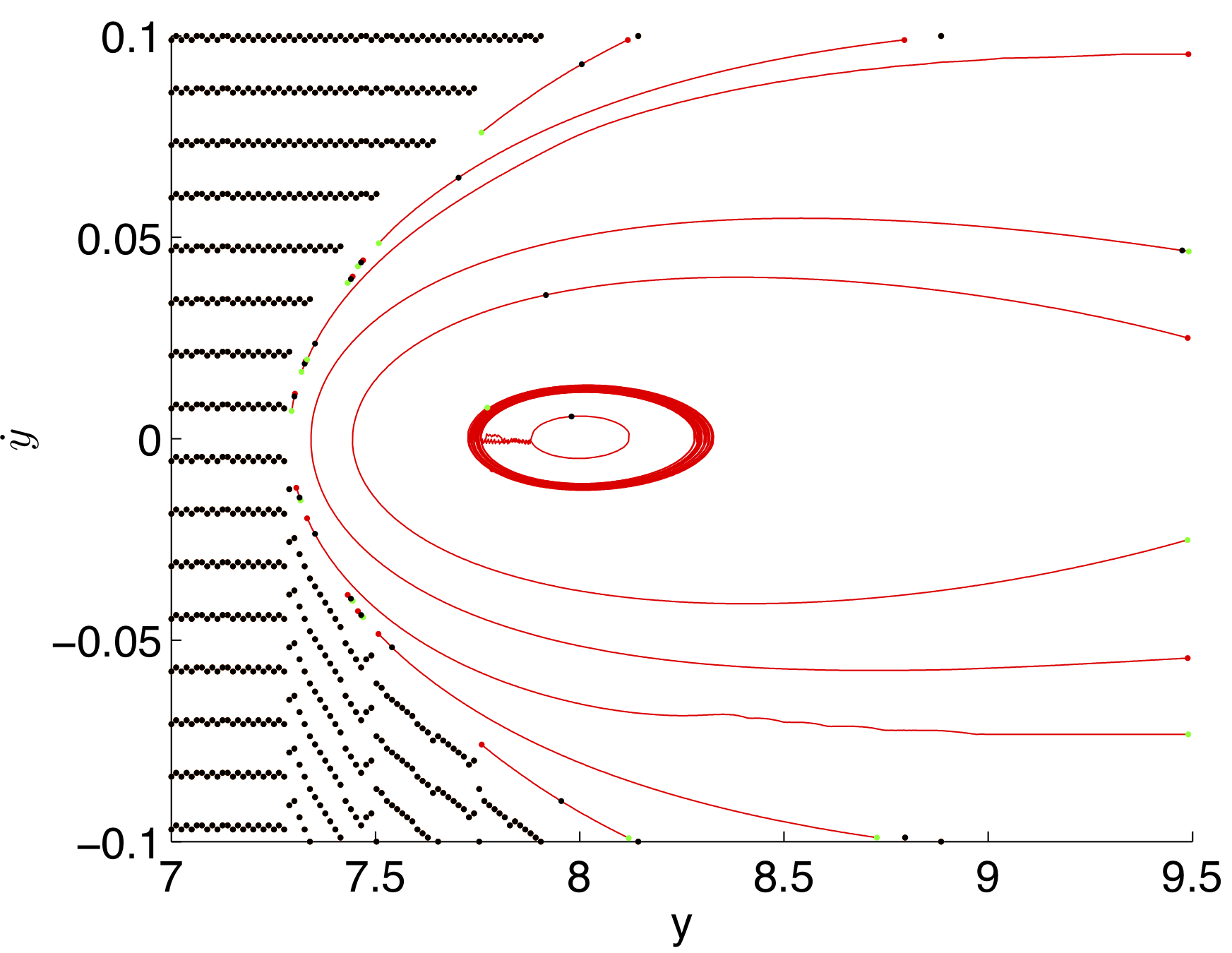} 
    \caption{FTLE (top) and strainlines (bottom) of the classical model for time $T=[0,505]$ in the spatial domain $[7,9.5] \times [-0.1,0.1]$.}
  \label{fig:strainprec505rangoamplio}
\end{figure}

For the integration time of $T=[0,505]$, for which we obtain good accuracy, we now increase the spatial domain $\Gamma$ to $[7,9.5] \times [-0.1,0.1]$ (Fig.~\ref{fig:strainprec505rangoamplio}).
The left part of both plots, in dark
blue in the FTLE field, shows the region of forbidden motion, where the black
dots indicate starting points in the computation of LCS. In the FTLE field as
well as in the strainlines, we observe structures within the fixed energy level
reflecting the dynamics of the system, corresponding to the heteroclinic orbits
and probably to further features, such as intersections of the parametrised
surface with other invariant manifolds.

\begin{figure}
\centering
    \includegraphics[width=0.32\textwidth]{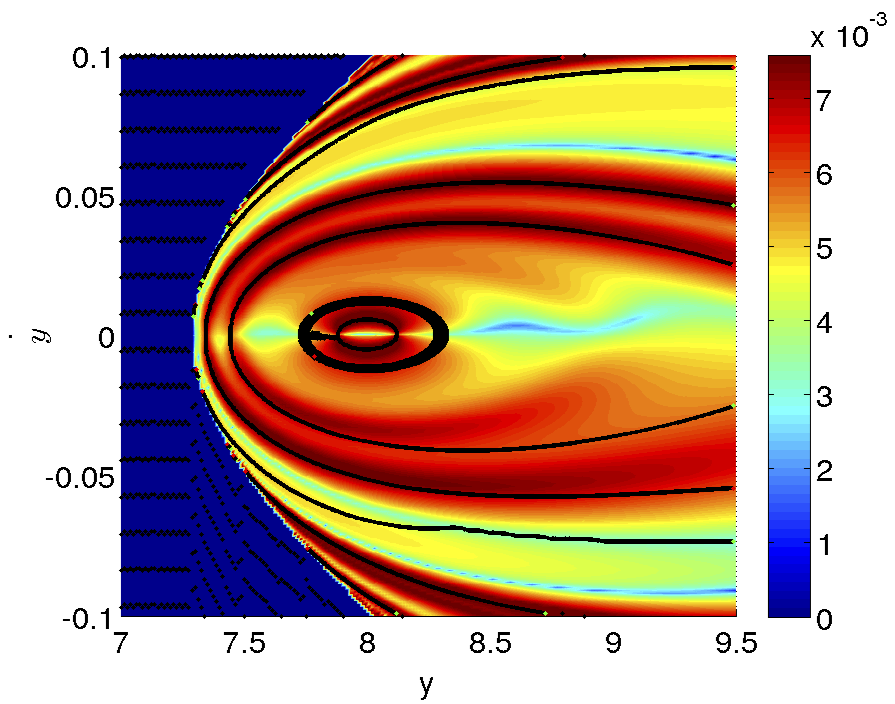} 
    \includegraphics[width=0.32\textwidth]{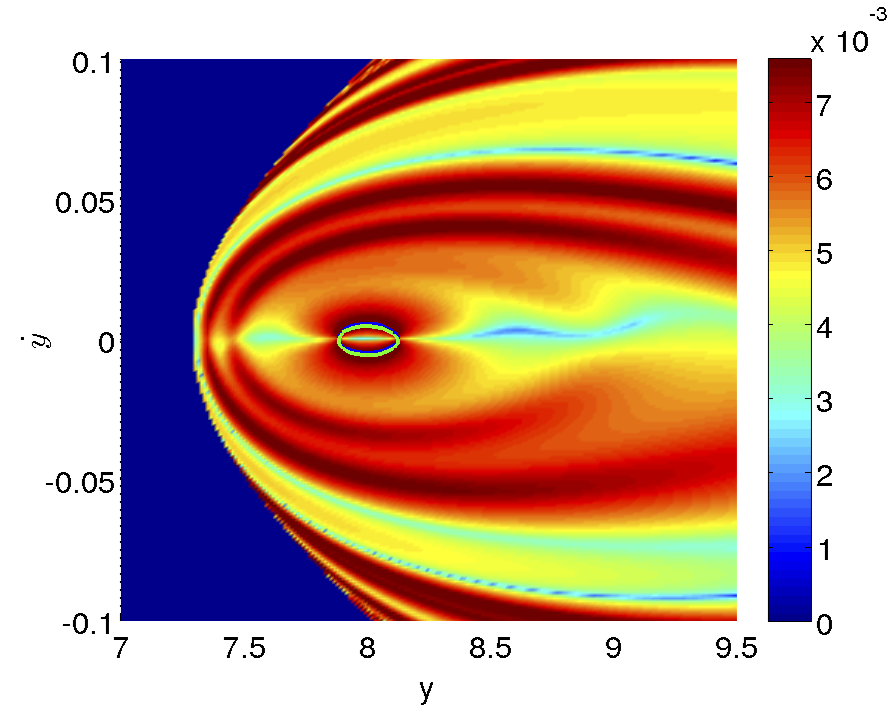} 
    \includegraphics[width=0.32\textwidth]{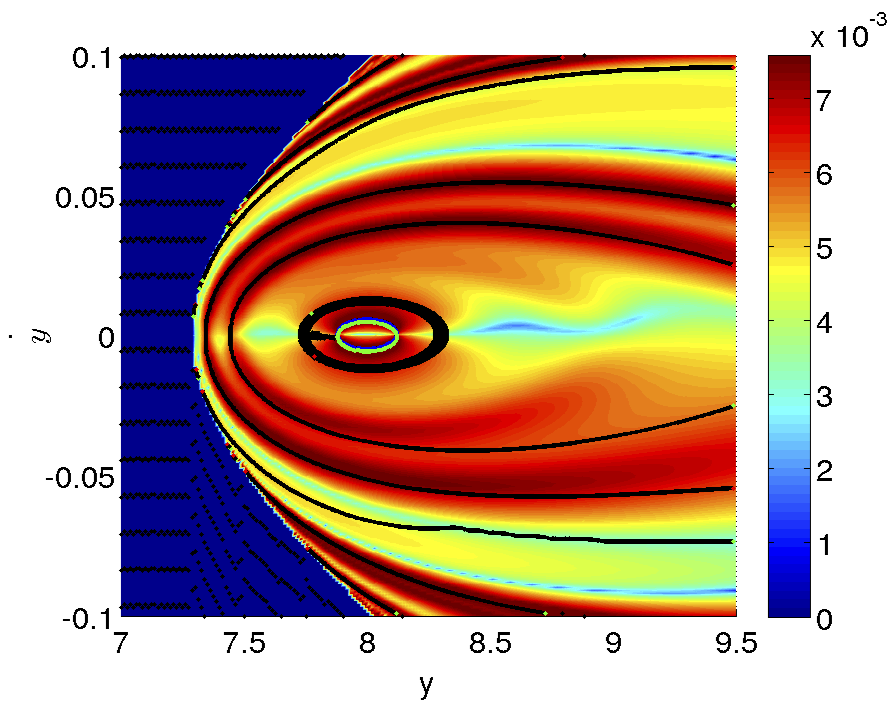} 
    \caption{Classical model for time $T=[0,505]$ in the spatial domain $[7,9.5] \times [-0.1,0.1]$. Top: FTLE field and strainlines in black. Center: FTLE field and heteroclinic orbits ($W_{\gamma_1}^{s,1}$ in blue, $W_{\gamma_2}^{u,1}$ in green). Bottom: FTLE field, strainlines (in black) and heteroclinic orbits (in blue and green).}
  \label{fig:strainheterprec505rangoamplio}
\end{figure}

Figure~\ref{fig:strainheterprec505rangoamplio} represents the superposition of the FTLE field and the strainlines (top panel), the FTLE field and the heteroclinic orbits (center panel) and the three elements in the bottom panel. The heteroclinic orbits correspond to the central closed curve, whereas the rest of the main ridges of the FTLE field are followed by the strainlines. This suggests that there are stable manifolds with the same level of energy associated to other structures, which are easily captured by the main ridges of the FTLE field and the corresponding strainlines.

\subsection{Lagrangian Coherent Structures in the non-autonomous case}

In the autonomous problem we observed that the FTLE field and its
corresponding strainlines are good indicators, analogue to the stable 
invariant manifolds. However the main purpose of Lagrangian Coherent Structures 
is to study the dynamics of non-autonomous problems where the computation of 
invariant manifolds is much more complex or when the basic structure associated
to the invariant manifold (e.g. equilibrium point or periodic orbit) does not
exist and so, the invariant manifolds are not defined. For these cases the 
computation of LCS still remains simple and valid, describing pretty well the 
organization of the motion moreover they can be seen as an extension of the 
invariant manifolds for the autonomous problems.
With this idea we transform our precessing model into a
non-autonomous model to observe the behaviour of its dynamics. 

According to \citet{Widrow,Manos} (among others) a parameter which is dependent on time in galaxy models is its pattern speed. The decrease in the pattern speed over time has been observed in galaxies, due to the transfer of angular momentum from the bar to the other components of the galaxy. So in this section, we consider a pattern speed depending on time in the precessing model, $\tOm = \mathbf{\Omega_p}(t)$, and this dependence introduces changes to the equations of the model. The vectorial form of the equations of motion in the rotating reference frame is now,
\begin{equation}\label{eq:motion_noaut}
{\mathbf{\ddot{r}=-\nabla \phi}} -2\mathbf{(\tOm \times \dot{r})}-\mathbf{\tOm \times (\tOm \times r)}-\tOmd \times \mathbf{r},
\end{equation}
where the term $-\tOmd \times \mathbf{r}$ is the inertial force of rotation~\citep[see][for further details]{bin08}. 
 
Taking $\tOm$ as in \citet{Warps}, but now with $\widetilde{\Omega}=\widetilde{\Omega}(t)$ depending on time,
\begin{equation}
\tOm=(-\widetilde{\Omega} \sin(\varepsilon),0,\widetilde{\Omega} \cos(\varepsilon)),
\label{eqn:angvelcte_noaut} 
\end{equation}
the new equations of motion are given by

\begin{equation}
 \left\lbrace
 \begin{array}{l}
  \ddot{x} = 2\widetilde{\Omega} \cos(\varepsilon) \dot{y} + \widetilde{\Omega}^2 \cos^2(\varepsilon) x + \widetilde{\Omega}^2 \sin(\varepsilon)\cos(\varepsilon)z+\dot{\widetilde{\Omega}}\cos(\varepsilon)y - \phi_{x} \\
  \ddot{y} = -2\widetilde{\Omega} \cos(\varepsilon)\dot{x} -2\widetilde{\Omega} \sin(\varepsilon)\dot{z} +\widetilde{\Omega}^2 y-\dot{\widetilde{\Omega}}\cos(\varepsilon)x-\dot{\widetilde{\Omega}}\sin(\varepsilon)z - \phi_{y} \\
  \ddot{z} = 2\widetilde{\Omega} \sin(\varepsilon)\dot{y} + \widetilde{\Omega}^2 \sin(\varepsilon) \cos(\varepsilon) x + \widetilde{\Omega}^2 \sin^2(\varepsilon) z+\dot{\widetilde{\Omega}}\sin(\varepsilon)y - \phi_{z}.\\
 \end{array}
 \right.
\label{eqn:systmodel_noaut}
\end{equation}

Let us remark that this system is non-autonomous, it has no integrals of
motion, and in particular it does not preserve energy, so there are no
constant-energy surfaces as in the autonomous precessing model. However in 
order to
parametrise a surface to compute the FTLE field and the strainlines, we
can consider the same conditions as in the previous case, taking the Jacobi
constant function of the autonomous problem, Eq.~(\ref{eqn:jacobicte}) with $\Omega=\widetilde{\Omega}$.

\begin{figure}
  \centering
    \includegraphics[width=0.31\textwidth]{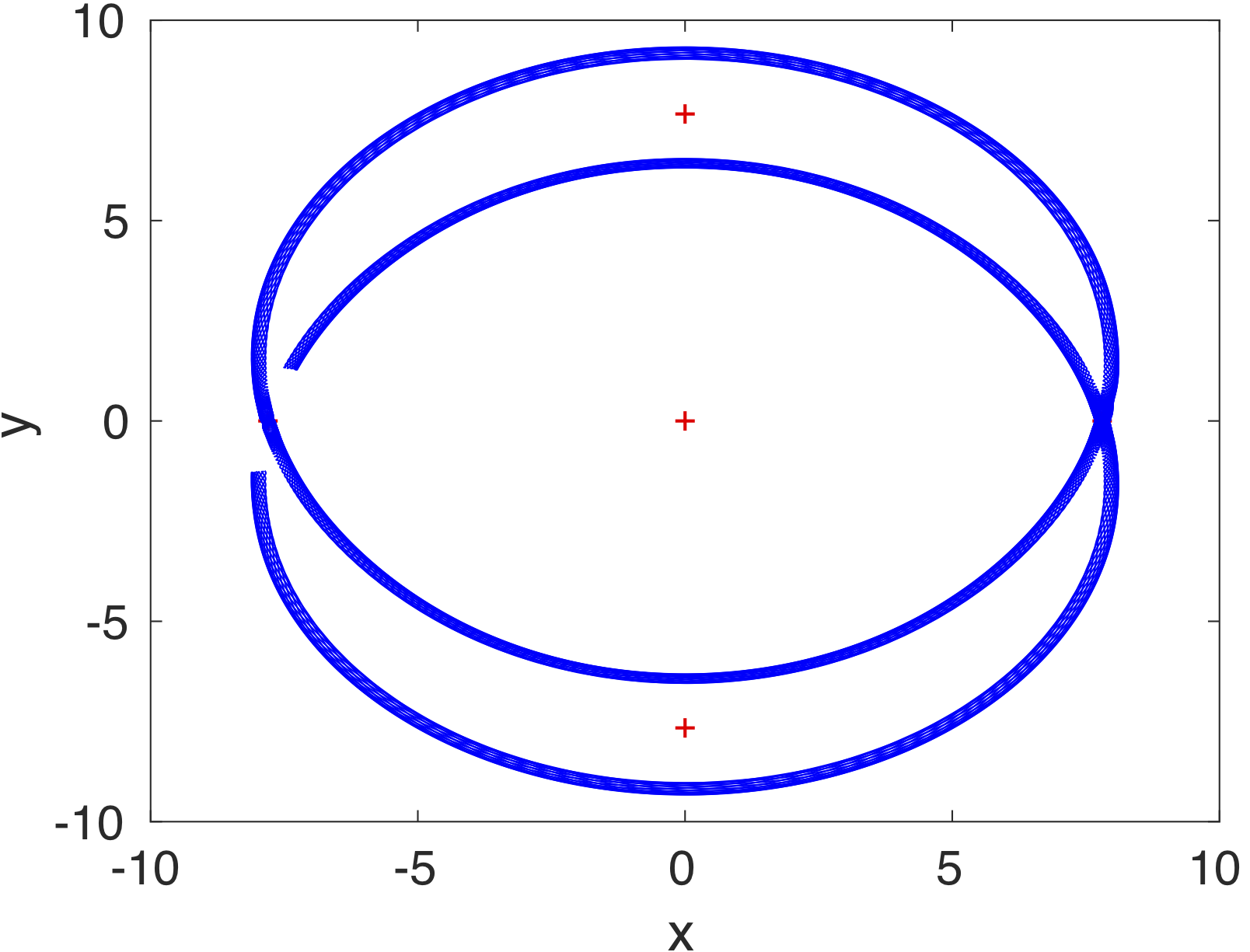}
  \caption{Unstable invariant manifolds (blue) and Lagrange points (red) of the model with mass bar
$GM_b=0.1$, tilt angle $\varepsilon=0$ and $\Omega=0.04$ in the $xy$ plane.}
\label{fig:peqOmega04}
\end{figure}

The parameters taken for the bar and disc are the same as previously, but now
the pattern speed varies linearly from $\widetilde{\Omega}_{t_0}=0.05$ to
$\widetilde{\Omega}_{t_f}=0.04$ for $[t_0,t_f]=[0,1500]$ time units, i.e. with
slope $\dot{\widetilde{\Omega}}=-\frac{2}{3}\cdot10^{-5}$. In the autonomous case, variations of pattern speed for $\Omega=0.05$ (shown in bottom panel of Fig.~\ref{fig:peq}) to $\Omega=0.04$ (shown in Fig.~\ref{fig:peqOmega04}) causes an increase of the period and the radius of the point $L_1$ by a factor of $1.2$. The effect in the galaxy is to open the internal ring. 

The decreasing slope of the pattern speed over time is in agreement to the behaviour observed in N-body simulations, due to the transfer of angular momentum from the bar to the other components of the galaxy \citep[e.g.][]{Widrow} In addition, the
starting ``energy'' level continues to be $C_{L_1}+\delta$, where $C_{L_1}$ is the
Jacobi constant for the equilibrium point $L_1$ in the autonomous problem. The
selected spatial domain for the integration is $\Gamma =
[7.5,8.7]\times[-0.02,0.02]$.  

\begin{figure}
    \includegraphics[width=0.45\textwidth]{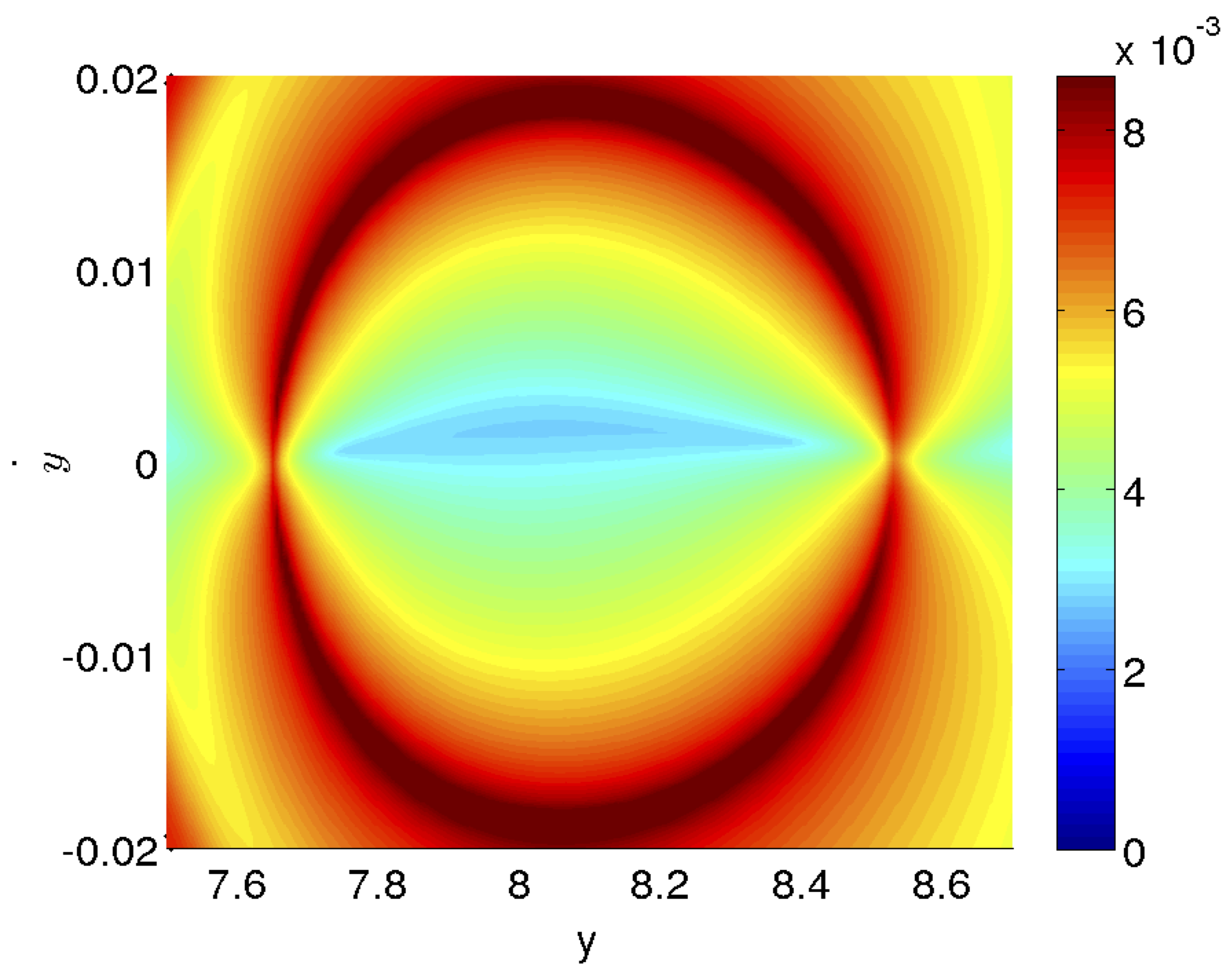} 
    \includegraphics[width=0.45\textwidth]{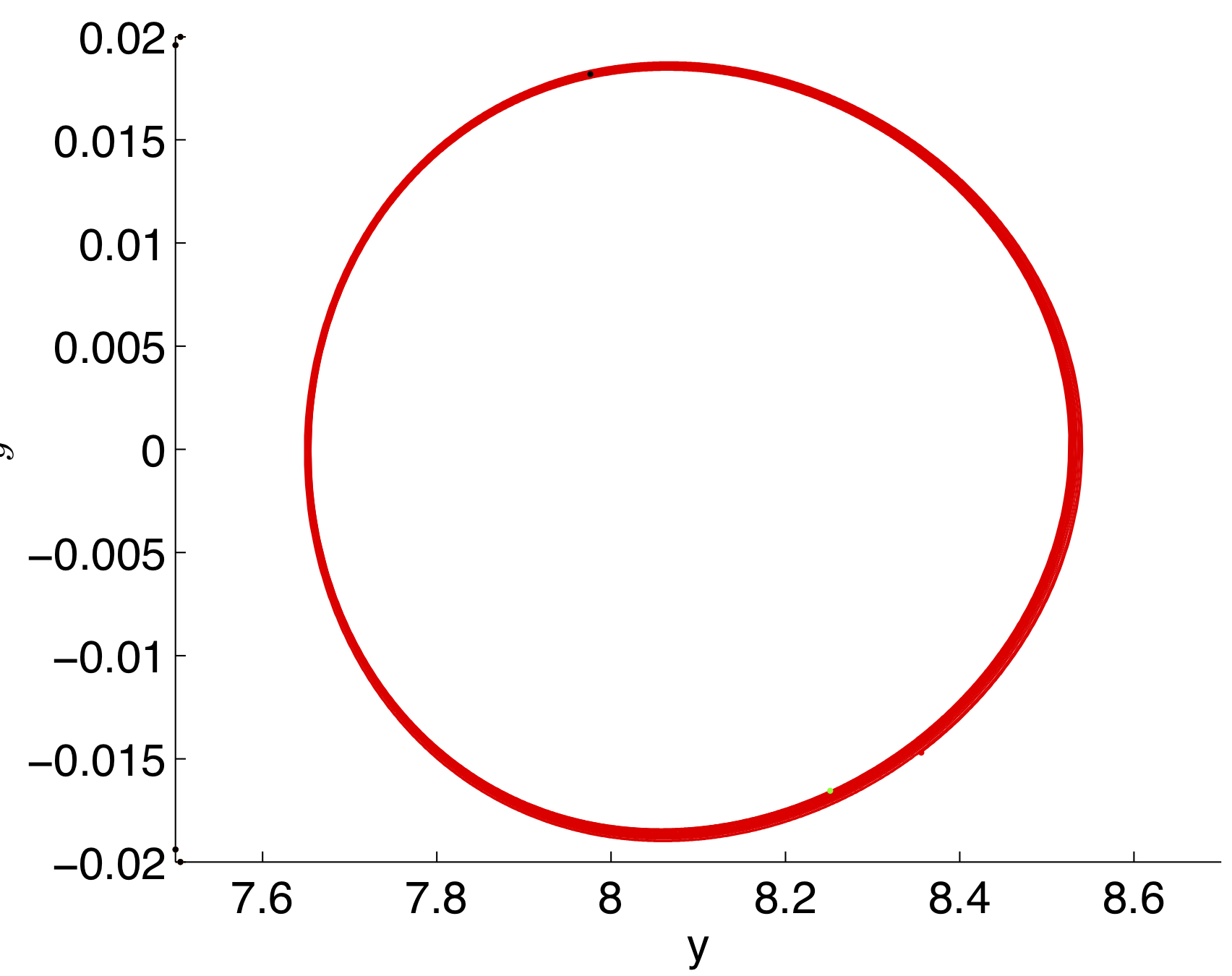} 
    \caption{FTLE (top) and strainlines (bottom) of the non-autonomous classical model with tilt angle $\varepsilon=0$ for time $T=[0,505]$.}
  \label{fig:strainprec505_noaut}
\end{figure}

\begin{figure}
\centering
    \includegraphics[width=0.24\textwidth]{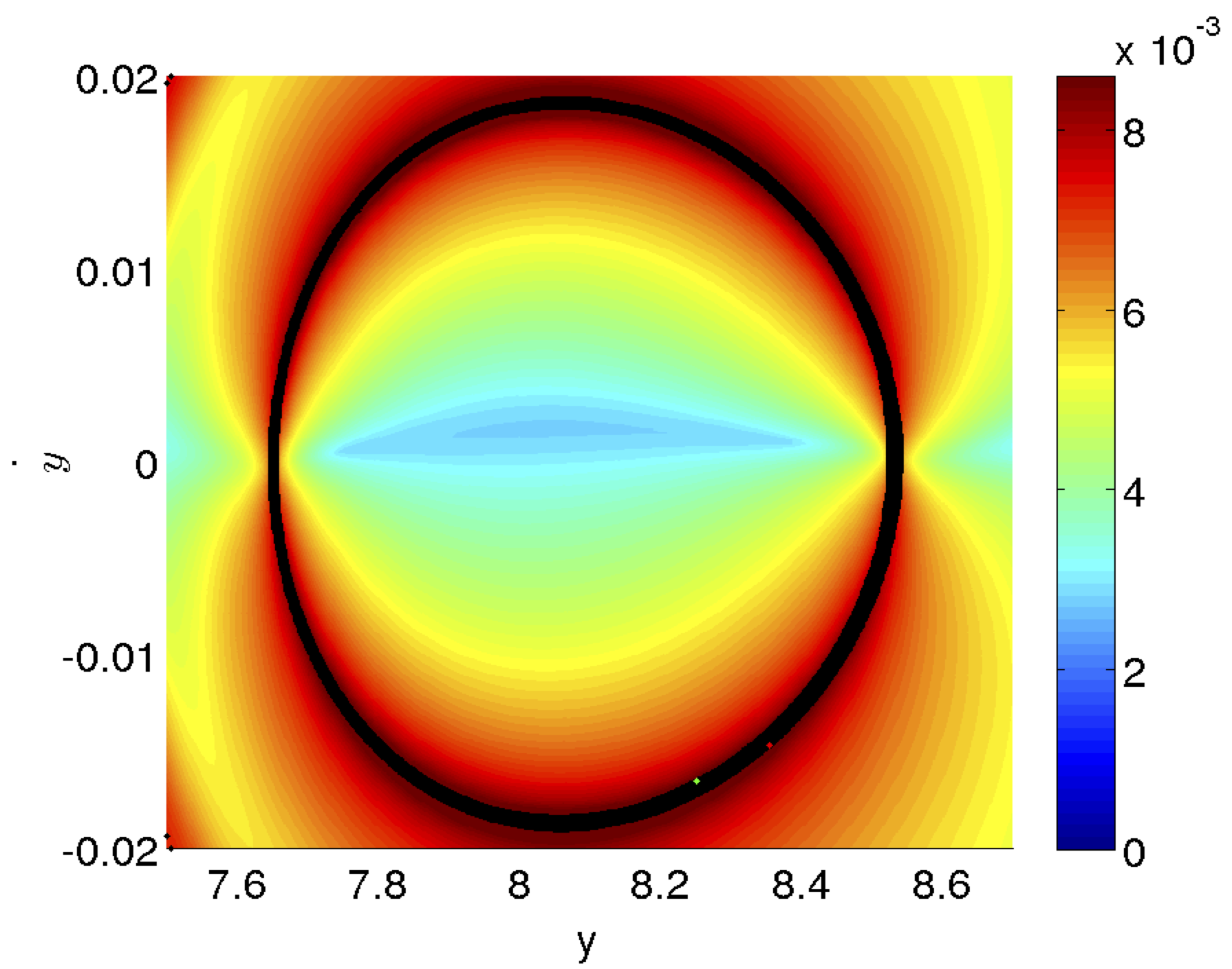} 
    \includegraphics[width=0.24\textwidth]{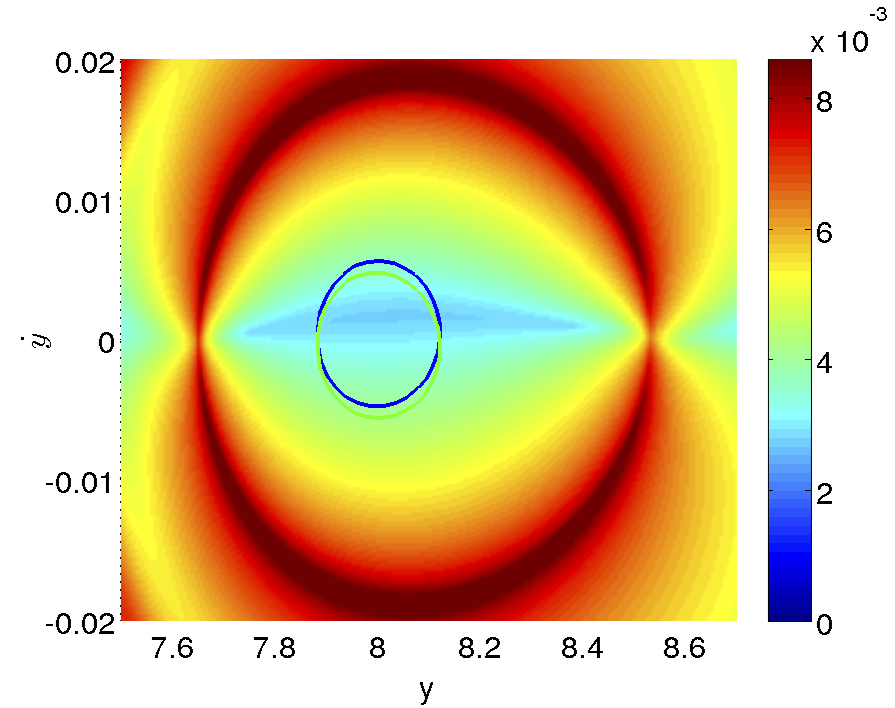} 
    \includegraphics[width=0.24\textwidth]{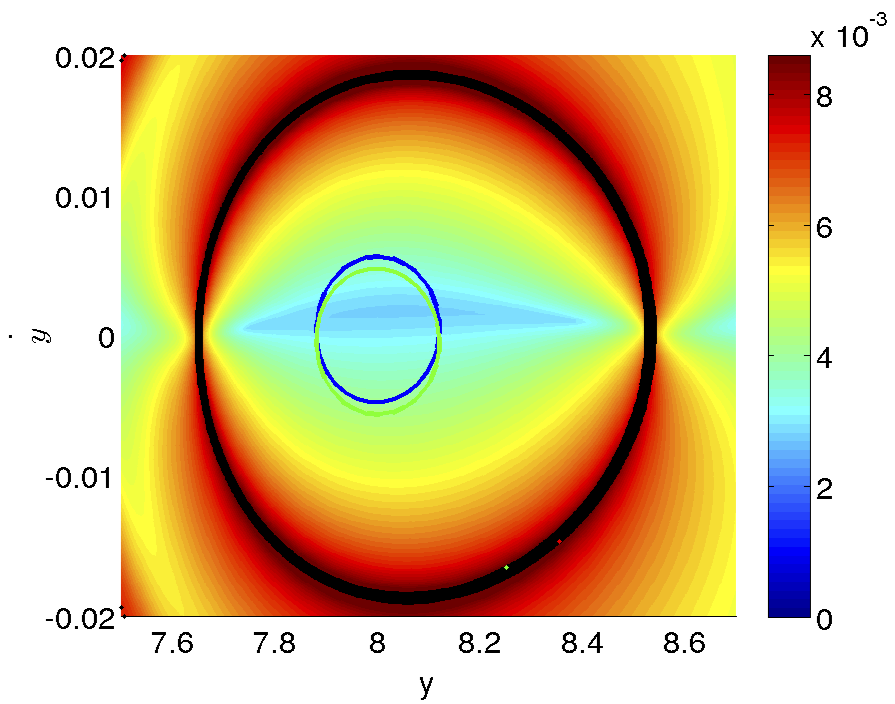} 
    \includegraphics[width=0.24\textwidth]{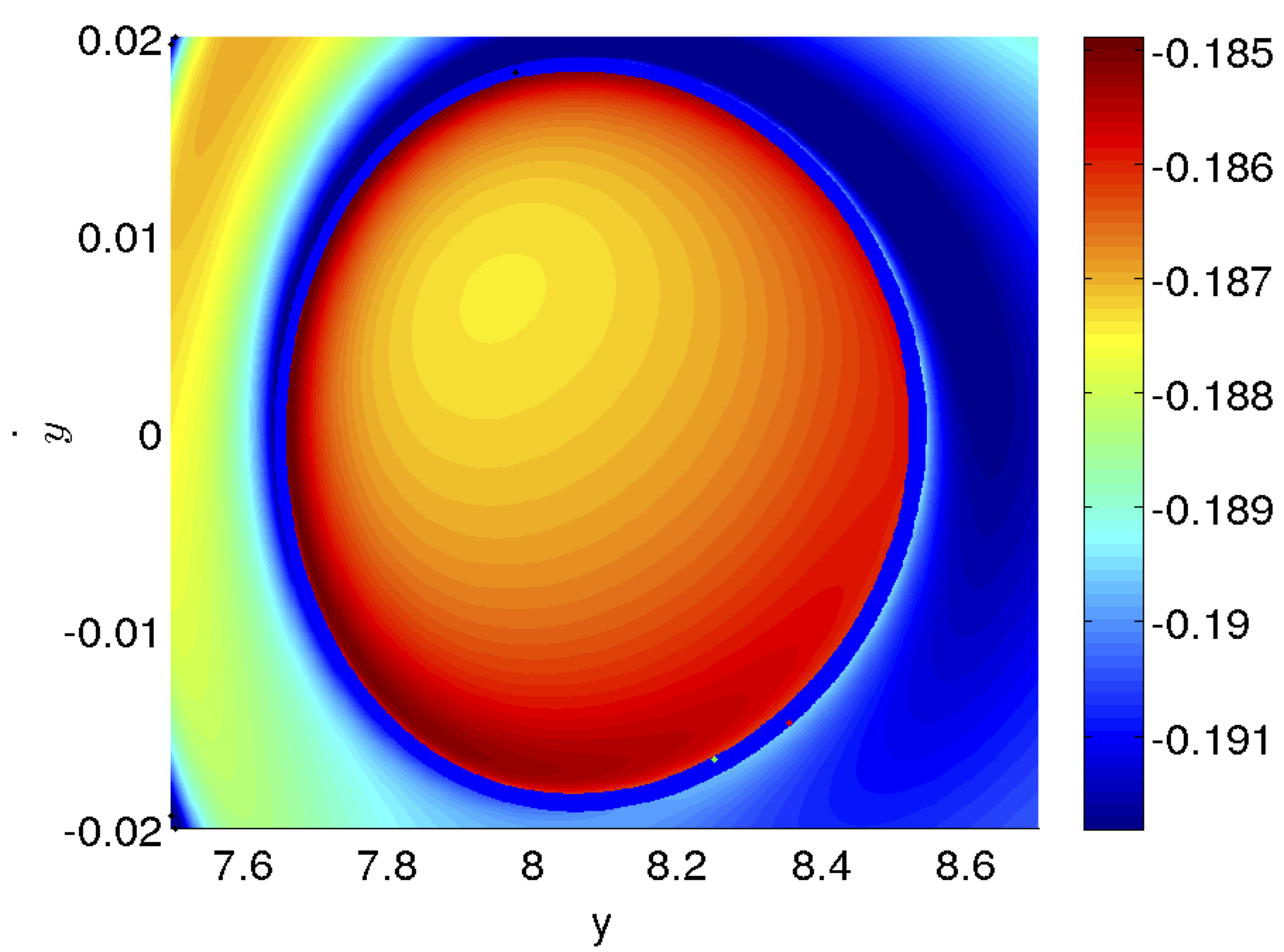} 
    \caption{Non-autonomous classical model for time $T=[0,505]$. Top left: FTLE field and strainlines in black. Top right: FTLE field and heteroclinic orbits of the autonomous classical model ($W_{\gamma_1}^{s,1}$ in blue, $W_{\gamma_2}^{u,1}$ in green). Bottom left: FTLE field, strainlines (in black) and heteroclinic orbits of the autonomous classical model (in blue and green). Bottom right: Energy at the endpoint of each orbit.}
  \label{fig:strainheterprec505_noaut}
\end{figure}

In order to compare the new results with the ones obtained for the autonomous
classical model, we take the same integration times. The first computation of
LCS is for the time interval $T=[0,505]$ (Fig.~\ref{fig:strainprec505_noaut}).
We observe that although the problem is non-autonomous, the shape of the main
ridges of the FTLE field and the strainlines continues being that of a closed
curve, and that the strainlines still follow the main ridges of the FTLE field.
But, if we superimpose the cuts $W_{\gamma_1}^{s,1}$ and $W_{\gamma_2}^{u,1}$
of the invariant manifolds of the autonomous model, we observe that
the widths of the FTLE field and strainlines have increased
(Fig.~\ref{fig:strainheterprec505_noaut}). Since the integration time is the
same as the one taken in the autonomous problem
(Fig.~\ref{fig:strainheterprec505}), the different range in the spatial domain
is due to a variation of energy when integrating the initial conditions. This
variation takes place because the function chosen to parametrise the surface
of initial conditions is not an integral of motion of the non-autonomous
model, and therefore the ``energy'' of the system given by this 
function in fact changes, as we
observe in the bottom right panel of Fig.~\ref{fig:strainheterprec505_noaut}.
Let us point out that in this figure the strainlines are associated with an
abrupt change of energy. 

\begin{figure}
    \includegraphics[width=0.45\textwidth]{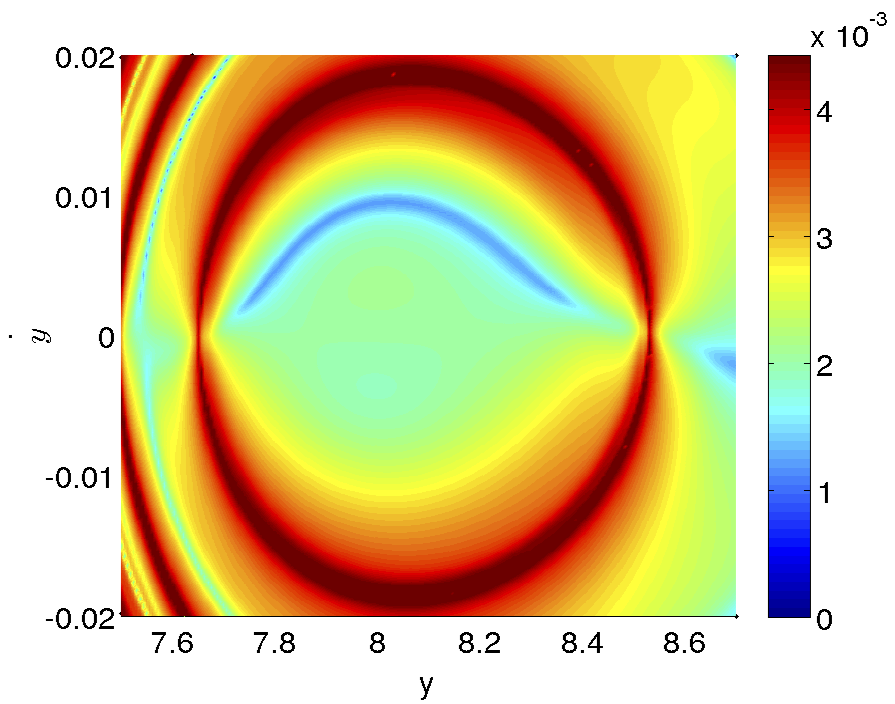} 
    \includegraphics[width=0.45\textwidth]{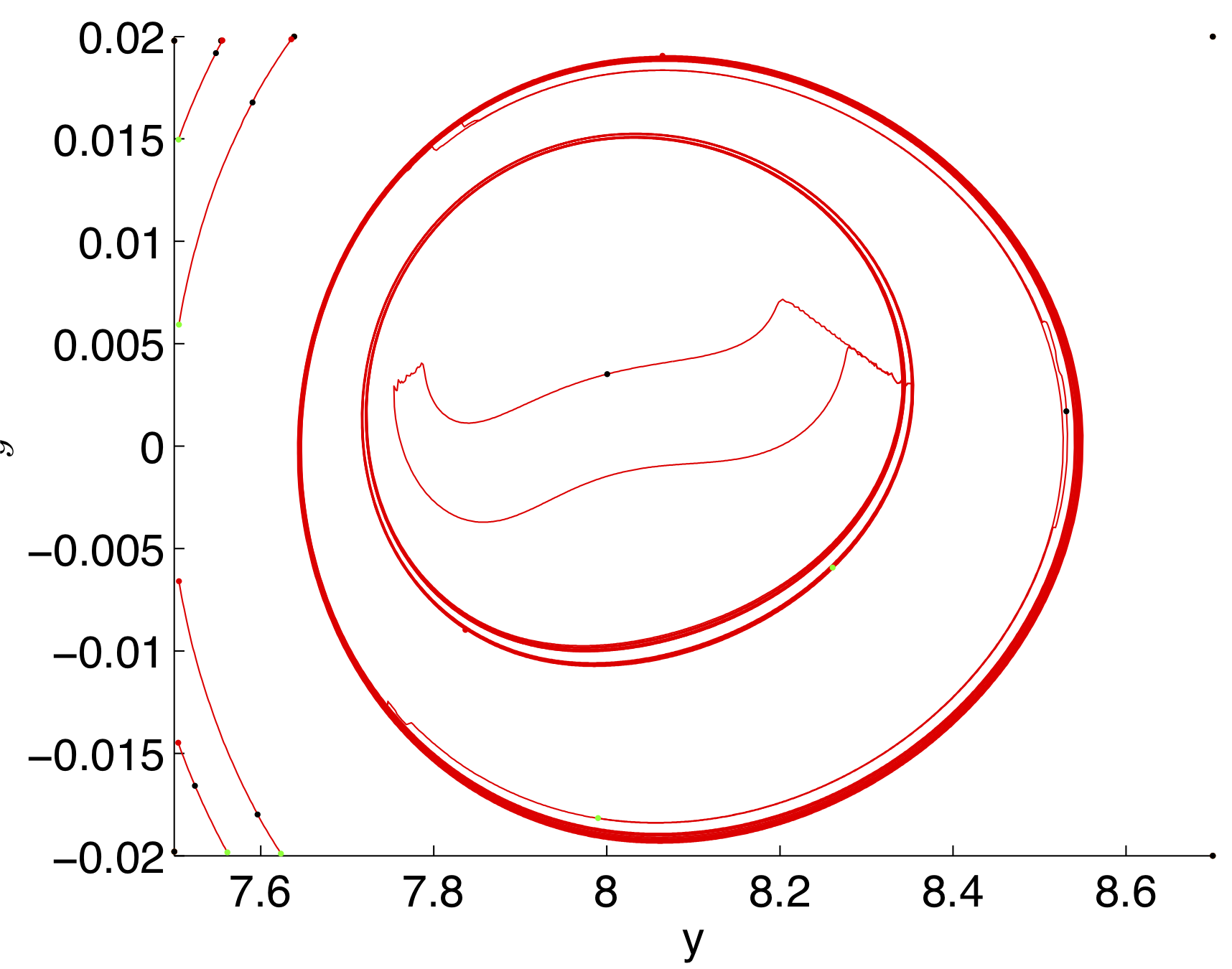} 
    \caption{FTLE (top) and strainlines (bottom) of the non-autonomous classical model for time $T=[0,1000]$.}
  \label{fig:strainprec1000_noaut}
\end{figure}

Figure~\ref{fig:strainprec1000_noaut} displays the FTLE field and the strainlines for an integration time of $T=[0,1000]$, and the same spatial domain as in the previous integration for time $T=[0,505]$ of the non-autonomous problem. While in the integration of the autonomous classical model both FTLE ridges and strainlines became blurred and separated from heteroclinic orbits when the integration time increased to $1000$, here they suffer a much smaller deformation compared to their position for integration time $T=[0,505]$.  

\begin{figure}
\centering
    \includegraphics[width=0.45\textwidth]{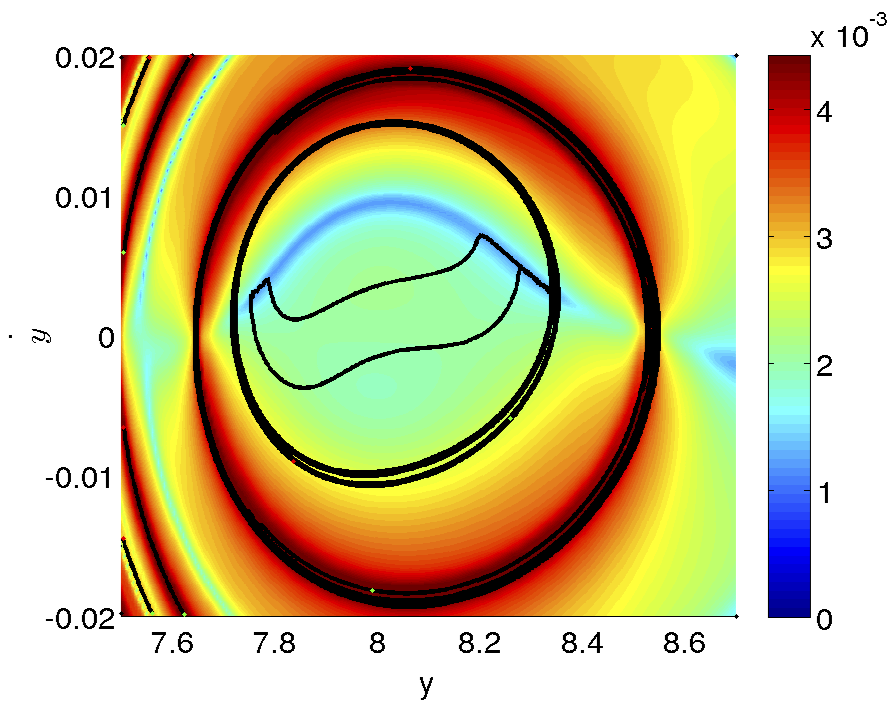} 
    \includegraphics[width=0.45\textwidth]{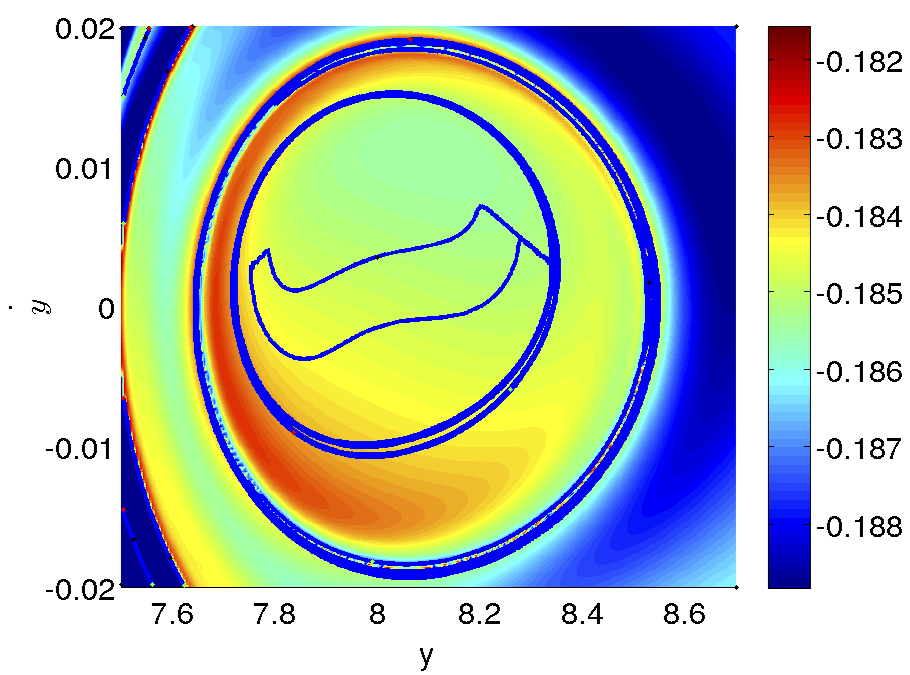} 
    \caption{Non-autonomous classical model for time $T=[0,1000]$. Top: FTLE field and strainlines in black. Bottom: Energy at the endpoint of each orbit.}
  \label{fig:strainftleprec1000_noaut}
\end{figure}

Figure~\ref{fig:strainftleprec1000_noaut} presents the strainlines overimposed on the FTLE field (top panel) and the final value of the energy in each orbit (bottom panel). Comparing the results with those of integration time $T=[0,505]$, we notice that an FTLE ridge-cum-strainline appears for values of $y$ up to $7.7$. The main strainlines follow the ridges of the FTLE field, but there is a secondary ring of central strainlines inside the main FTLE ridge that does not correspond to any feature of the FTLE field. In the bottom panel we observe that the main strainlines coincide with an abrupt variation of the final energy level of each orbit, and the central strainlines are placed over a smoother variation of the final energy level. An isolated central strainline diverging from the inner ring seems an artifact introduced by inaccuracies accumulated over the long integration time.

\begin{figure}
\centering
    \includegraphics[width=0.45\textwidth]{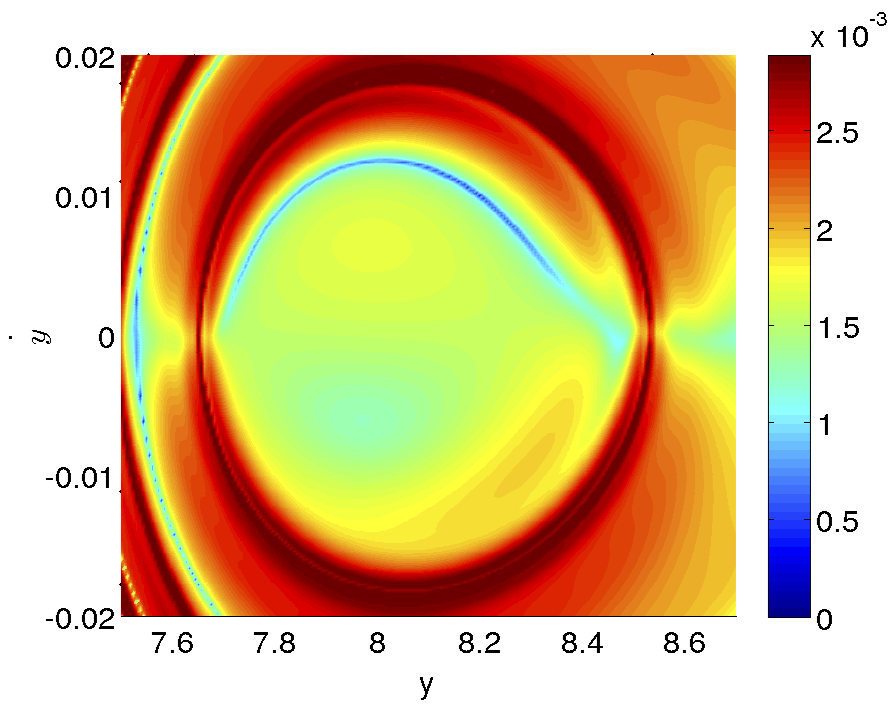} 
    \includegraphics[width=0.45\textwidth]{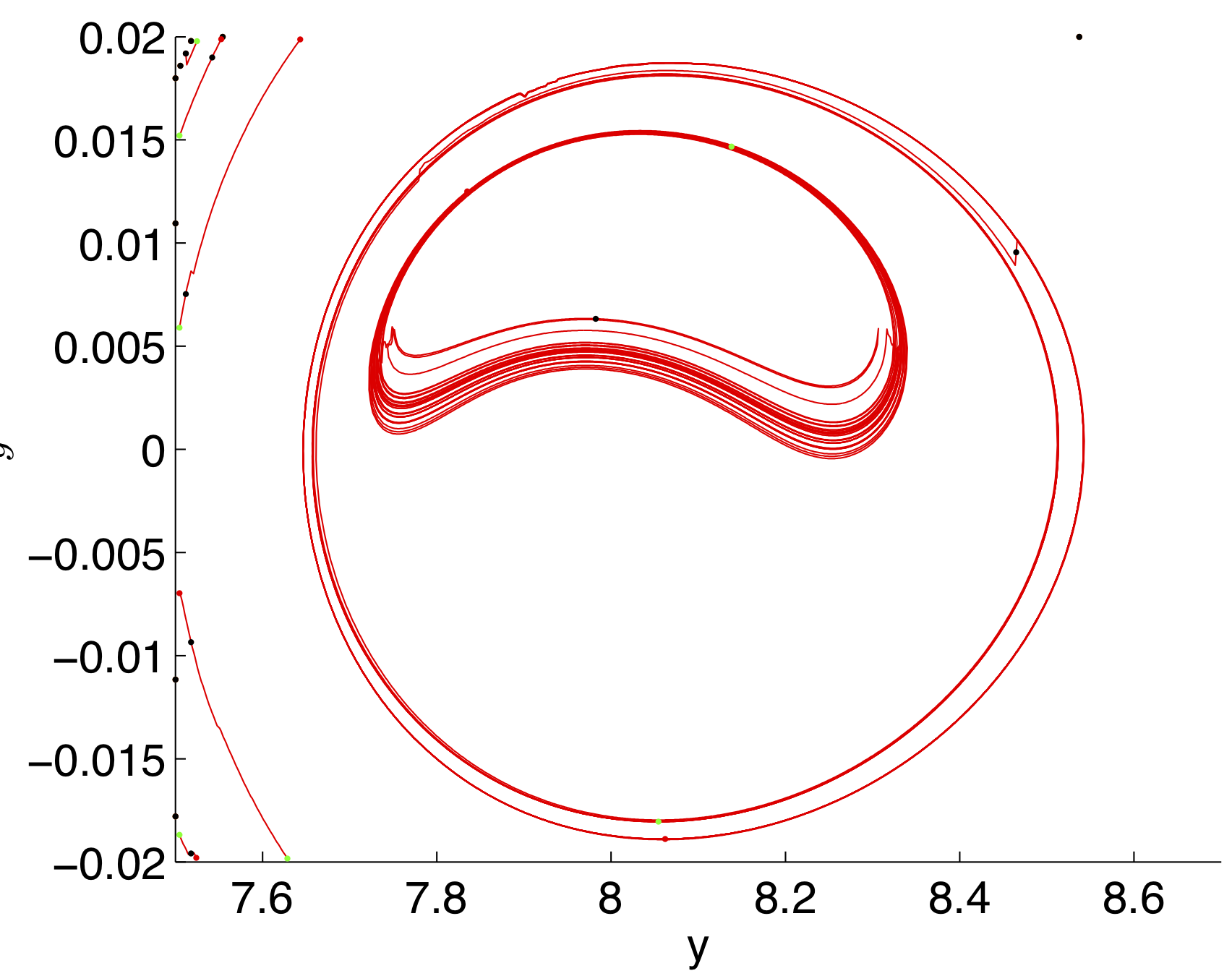} 
    \caption{FTLE (top) and strainlines (bottom) of the non-autonomous precessing model with tilt angle $\varepsilon=0$ for time $T=[0,1570]$.}
  \label{fig:strainprec1570_noaut}
\end{figure}

Figure~\ref{fig:strainprec1570_noaut} shows the FTLE field and the strainlines for an interval of integration of $T=[0,1570]$. The distribution of values of the FTLE field is very similar to that of the interval of integration $T=[0,1000]$. Regarding the strainlines, they cover the ridges of the FTLE field and there is a further inner ring of strainlines inside the main ridge of the FTLE field following a lesser ridge that borders areas of decrease of the FTLE field.

\begin{figure}
\centering
    \includegraphics[width=0.45\textwidth]{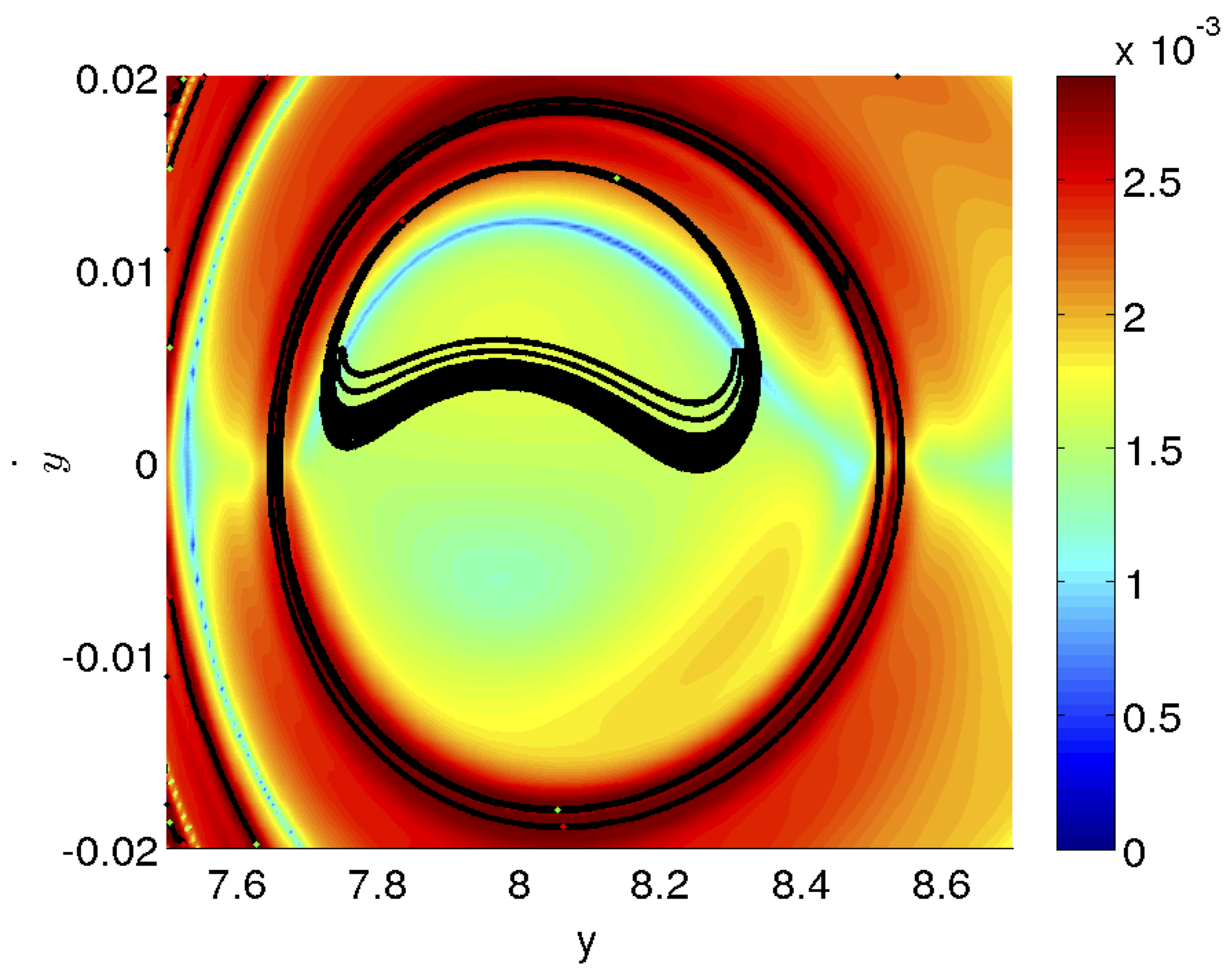} 
    \includegraphics[width=0.45\textwidth]{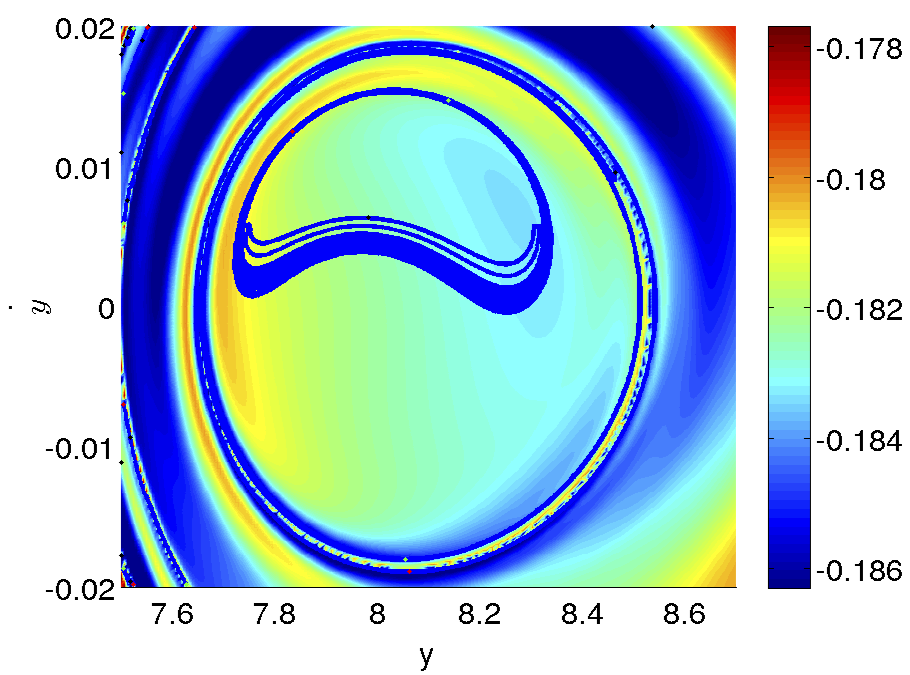} 
    \caption{Non-autonomous precessing model with tilt angle $\varepsilon=0$ for time $T=[0,1570]$. Top: FTLE field and strainlines in black. Bottom: Energy at the endpoint of each orbit.}
  \label{fig:strainftleprec1570_noaut}
\end{figure}

Figure~\ref{fig:strainftleprec1570_noaut} illustrates that the behaviour of the strainlines in relation to the final energy of the orbits for integration time $T=[0,1570]$ is the same as for integration time $T=[0,1000]$. Therefore, the main rings of strainlines border on the steepest variations of energy level and the inner ring of strainlines also borders on a secondary area of variation of the energy level. 

\begin{figure}
\centering
    \includegraphics[width=0.24\textwidth]{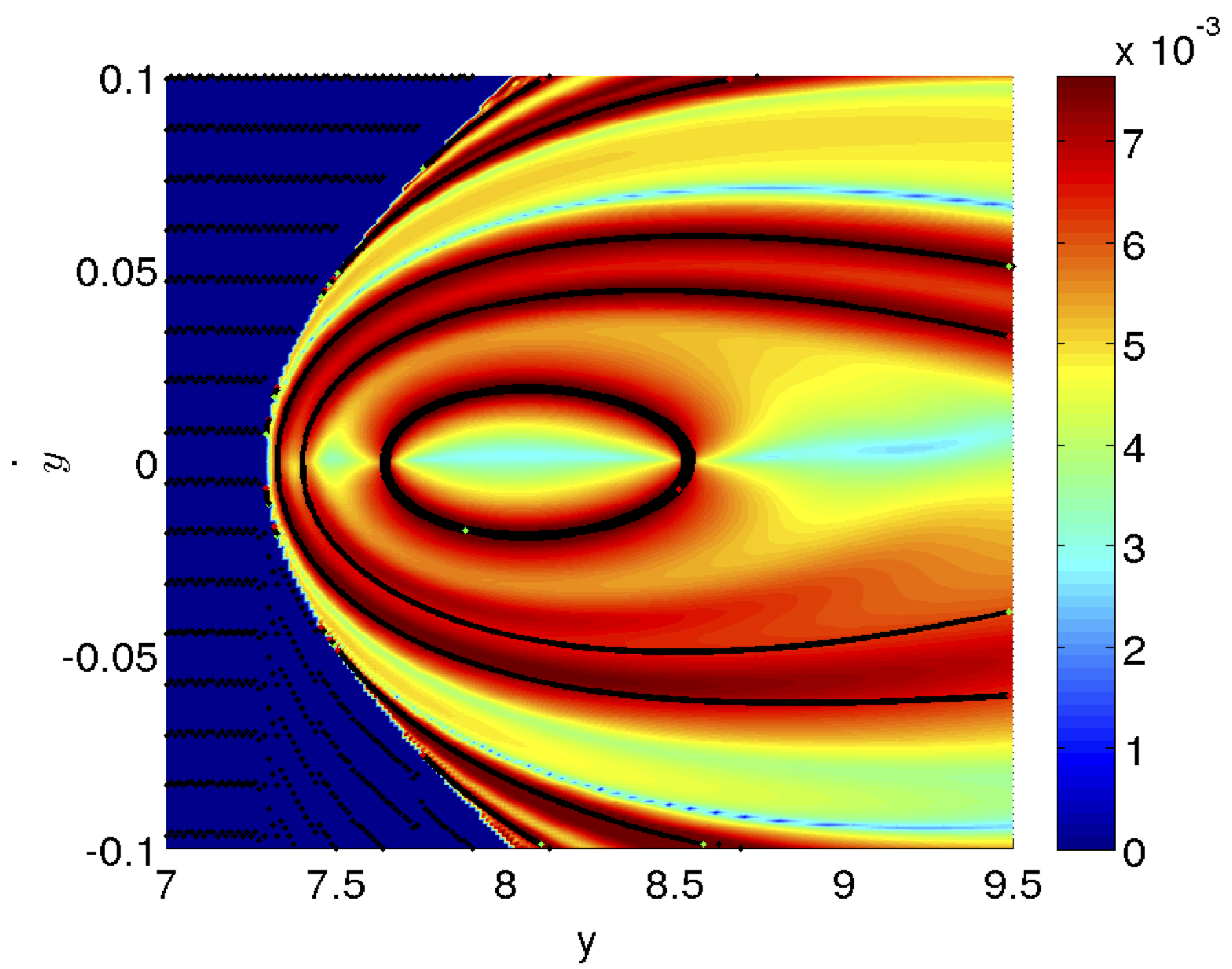} 
    \includegraphics[width=0.24\textwidth]{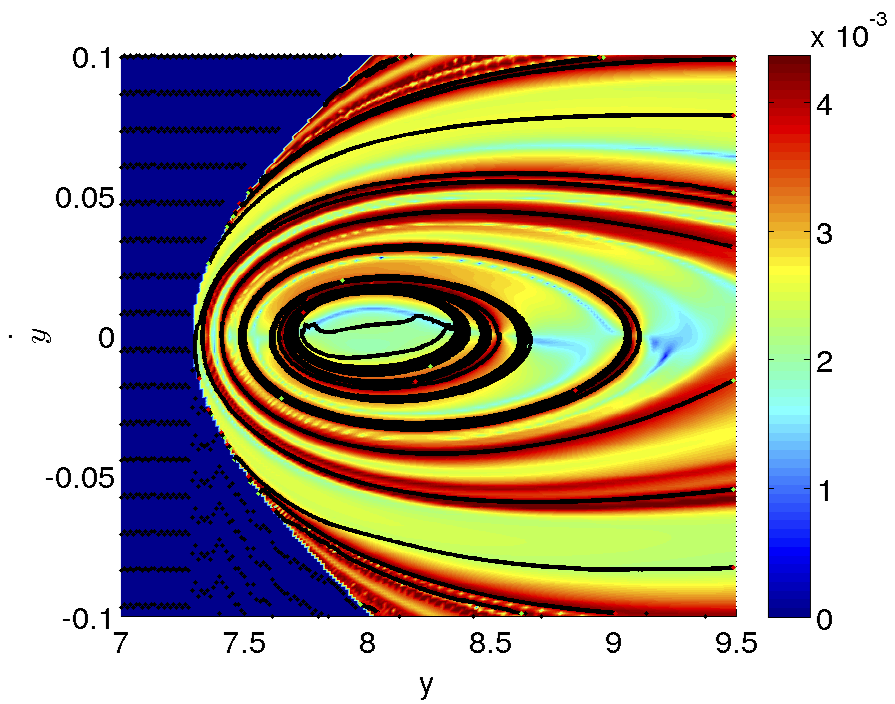}
    \includegraphics[width=0.24\textwidth]{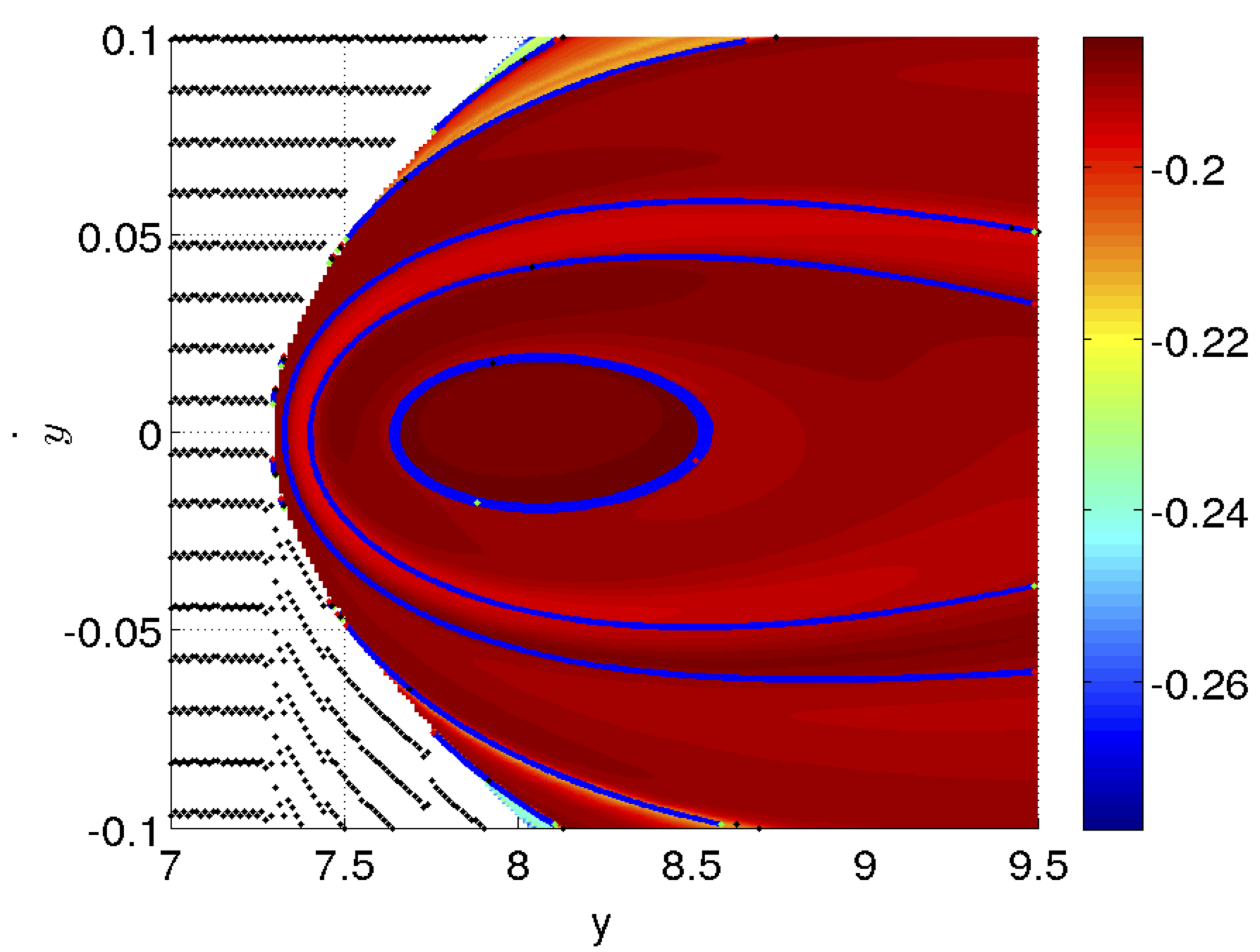} 
    \includegraphics[width=0.24\textwidth]{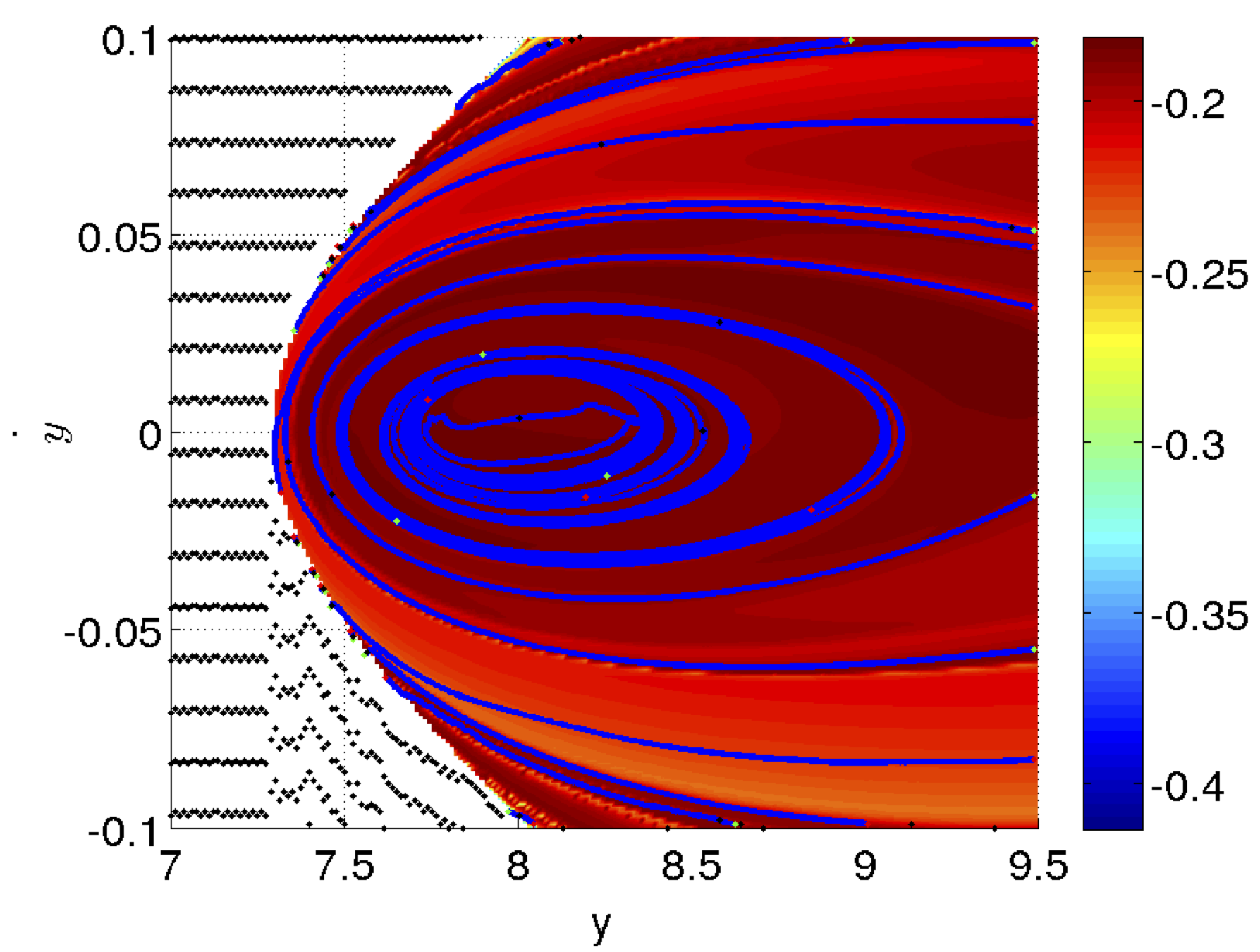}
    \caption{Non-autonomous precessing model with tilt angle $\varepsilon=0$ for time $T=[0,505]$ and $T=[0,1000]$. Top: FTLE field and strainlines in black for time $T=[0,505]$ (left) and $T=[0,1000]$ (right). Bottom: Energy at the endpoint of each orbit for time $T=[0,505]$ (left) and $T=[0,1000]$ (right), and strainlines (in blue).}
  \label{fig:strainftleprec505y1000_noaut}
\end{figure}

Finally Fig.~\ref{fig:strainftleprec505y1000_noaut} shows the strainlines, FTLE
field and final energy level for the orbits for a wider parametrised surface in
the spatial domain $(y,\dot{y})\in \Gamma=[7,9.5] \times [-0.1,0.1]$, for
integration times $T=[0,505]$ in the left column and $T=[0,1000]$ in the right
column. Looking at the FTLE field and the strainlines (top row) we see that
there are more ridges of the FTLE field and strainlines surrounding the main
one seen on the previous figures. A longer integration time leads to the
appearance of further features (FTLE ridges and strainlines) of this type. The
comparison of the final energy level of the orbits with the strainlines (bottom
row) establishes that the strainlines point out curves of steepest variation
for the final energy level of the orbits.  

To sum up the computations of this section, we can state that FTLE ridges are covered by strainlines, that there appear some further strainlines associated to secondary features of the FTLE field (typically abrupt descents of the field value) and that strainlines in both cases mark curves of steepest descents of the final energy level of the orbits. 

Our computations also show that the non-autonomous precessing model presents more stable FTLE field and strainlines than the autonomous version. This greater stability might be due to the energy-dissipative character of our non-autonomous model.

\section{Discussion and conclusions} \label{sec5:conclu}
One of the main concerns when applying the manifold theory to barred galactic
models is to know how it behaves when the system is under a certain secular
evolution. 
The main goal of this paper is to establish the behaviour of our model in its
non-autonomous version, in particular we want to assess whether an adiabatic
decrease of the bar pattern speed can change the global morphology of the
galaxy, as predicted by the invariant manifolds in the autonomous model. Since
invariant manifolds do not exist as such in the non-autonomous problem, we
introduce the LCS, and therefore, this paper deals with the study of the
non-autonomous version of the galactic model in 2 dimensions by means of LCS.
This is a recently developed theory to determine dynamical structures, either
attracting or repelling, reflecting the dynamics of the system in
non-autonomous problems. To apply this theory to our galactic model, we have
created our own LCS computation software, made as accurate and efficient as
possible, and capable of computing LCS in general 2-dimensional dynamical
systems and in parametrized surfaces in dynamical systems of any dimension. The
LCS are based on the singular values and vectors of the Jacobian of the flow,
equivalent to the eigenvalues and the eigenvectors of the Cauchy-Green tensor.
In particular, the Finite Time Lyapunov Exponent (FTLE) field is determined by
the main singular value, i.e. the norm, of the Jacobian of the flow. Roughly
speaking, these LCS give us the repulsion and attraction zones. In
2-dimensional domains, we obtain strainlines and stretchlines which are the
maximally repelling or attracting lines, respectively. Whereas for the
strainlines we have found that they give us relevant information about the
dynamics of the system, the strecthlines have turned out to be more
ill-conditioned and less connected to the global dynamics of the system. Let us
also remark that the LCS must be applied to ``smooth problems'', in the sense
that the problems have no abrupt changes in its behaviour, since the
computation of the flow Jacobian and, above all, of the stretchlines, is very
sensitive to sudden variations of the integrated vector field.

In order to better understand the information given by the FTLE field and the
LCS, we apply it first to the autonomous galactic model. By selecting a parametrized
surface given by the energy of the system in one of the equilibrium points
(L$_1$ or L$_2$), and initial conditions in the $(y,\dot{y})$ plane, we obtain
that the flow Jacobian norm field points to the zones where the stable
invariant manifold is placed, and the strainlines accurately overlap with this
stable manifold. As the integration time increases, the unavoidable build up of
error in the computation of the flow causes it to lose precision in the related
flow Jacobian norm field and therefore in the strainline computations, but
these still approximate the related stable manifolds. Moreover, the flow
Jacobian norm field and the LCS give information about other zones of maximal
repulsion placed in a bigger spatial domain, which seem to correspond to
invariant manifolds caused by other structures.

We then apply the LCS to the non-autonomous problem, although the energy
is not preserved anymore. Selecting the same time intervals of integration as
previously, we observe that the flow Jacobian norm field and the strainlines
continue remarking the zones of maximal repulsion, with a shape analogous to
the previous stable invariant manifold, but with a greater width. In contrast
with the autonomous case, in this non-autonomous galactic model the flow
Jacobian norm field and the strainlines are not distorted as the integration
time increases. This stronger stability seems to be a consequence of the
energy-dissipative character of the selected model. In addition, the other
zones of maximal repulsion found in the autonomous problem continue existing in
this time-dependent model, but now we observe that these zones are placed where
a steepest change on the energy of the system happens. 

From this study we can derive two main conclusions. First, the LCS strainlines of the non-autonomous galactic problem indicate zones of maximal repulsion in the domain and they are related to the stable manifolds of the galactic autonomous problem when this problem is integrated forward in time, or to its unstable manifolds when it is integrated backwards in time. Second, for a fixed set of input parameters defining the autonomous galactic problem, the invariant manifolds drive the motion of stars through the unstable periodic orbits giving the galaxy a certain morphology, namely ringed or spiral barred galaxy. When allowing a certain adiabatic secular evolution to the system, in this case, in the form of a slow decrease of the bar pattern speed with time, as seen from N-body simulations \cite[][e.g.]{Widrow}, the invariant manifolds prediction still holds, since the LCS and the strainlines generalize the dynamics predicted by the manifolds.

\section*{Acknowledgements}

This work started as part of the doctoral dissertation of P.S.M., supported by the Catalan PhD grants FI-AGAUR and FPU-UPC. J.J.M. thanks MINECO-FEDER (Spanish Ministry of Economy) for the grant MTM2015-65715-P, and the Catalan government for the grant 2017SGR1049. M.R.G. thanks MINECO for the grant ESP2016-80079-C2-1-R (MINECO/FEDER, UE) and MDM-2014-0369 of ICCUB (Unidad de Excelencia 'María de Maeztu').



%

\end{document}